\documentclass[useAMS,usenatbib,longnamesfirst,usedcolumn]{mn2e} 
\usepackage{epsfig}	
\usepackage{times}	
\title[A statistical analysis of the 2dXGS: Feedback]{A statistical analysis of the Two Dimensional \textit{XMM-Newton} Group Survey: The impact of feedback on group properties}
\author[R. Johnson, T. J. Ponman and A. Finoguenov]
       {Ria Johnson$^{1}$\thanks{E-mail: ria@star.sr.bham.ac.uk},
        Trevor J. Ponman$^{1}$, Alexis Finoguenov$^{2}$\\
        $^{1}$School of Physics and Astronomy, University of
        Birmingham, Edgbaston, Birmingham B15 2TT, UK\\
        $^{2}$Max-Planck-Institut f\"{u}r extraterrestrische Physik, 
        Giessenbachstra{\ss}e, 85748 Garching, Germany\\
       \\}
 \date{Accepted 2009 February 14. Received 2009 February 11;
      in original form 2009 January 09}

\pagerange{\pageref{firstpage}--\pageref{lastpage}}
\pubyear{2009}

\shortcites{RC3,degrandi04,peres98,voit03b,paturel03,napolitano05,sun08,finoguenov06,finoguenov07,machacek05,buote03,osullivan07,kochanek01,mccarthy08,mulchaey03,trinchieri03,trinchieri05,osullivan07b,jetha08,finoguenov02,mahdavi05,mahdavi00,jetha07,isobe90,voit02,moll07,humphrey06,magorrian98,baldi07,borgani08a,borgani08b}

\newcommand{\rmsub}[2]{\ensuremath{#1_{\mathrm{#2}}}} 
\newcommand{\Chandra}{\emph{Chandra}}
\newcommand{\Rosat}{\emph{ROSAT}}
\newcommand{\cm}{\ensuremath{\mbox{~cm}}}
\newcommand{\erg}{\ensuremath{\mbox{~erg}}}
\newcommand{\ergps}{\ensuremath{\erg \ps}}
\newcommand{\keV}{\ensuremath{\mbox{~keV}}}
\newcommand{\kpc}{\ensuremath{\mbox{~kpc}}}
\newcommand{\LX}{\rmsub{L}{X}}
\newcommand{\Msol}{\rmsub{M}{\odot}}
\newcommand{\ps}{\ensuremath{\s^{-1}}}
\newcommand{\s}{\ensuremath{\mbox{~s}}}
\newcommand{\XMM}{\emph{XMM-Newton}}
\newcommand{\RF}{\ensuremath{r_{500}}}

\newcommand{\rr}{\textsc{r}}
\newcommand{\D}{\ensuremath{D_{25}}}
\newcommand{\Hn}{\ensuremath{H_{0}}}

\begin{document}

\maketitle

\label{firstpage}

\begin{abstract}
  
 \noindent We have performed a statistical analysis of a sample of 28
 nearby galaxy groups derived primarily from the Two-Dimensional \XMM\
 Group Survey (2dXGS), in order to ascertain what factors drive the
 observed differences in group properties. We specifically focus on
 entropy and the role of feedback, and divide the sample into cool
 core (CC) and non cool core (NCC) systems. This is the first time the
 latter have been studied in detail in the group regime. We find the
 coolest groups to have steeper entropy profiles than the warmest
 systems, and find NCC groups to have higher central entropy and to
 exhibit more scatter than their CC counterparts. We investigate the
 entropy distribution of the gas in each system, and compare this to
 the expected theoretical distribution under the condition that
 non-gravitational processes are ignored. In all cases, the observed
 maximum entropy far exceeds that expected theoretically, and simple
 models for modifications of the theoretical entropy distribution
 perform poorly. A model which applies initial pre-heating through an
 entropy shift to match the high entropy behaviour of the observed
 profile, followed by radiative cooling, generally fails to
 match the low entropy behaviour, and only performs well when the
 difference between the maximum entropy of the observed and
 theoretical distributions is small. Successful feedback models
 need to work differentially to increase the entropy range in the gas,
 and we suggest two basic possibilities.

 We analyse the effects of
 feedback on the entropy distribution, finding systems with a high
 measure of `feedback impact' to typically reach higher entropy than
 their low feedback counterparts. The abundance profiles of
 high and low feedback systems are comparable over the majority of the
 radial range, but the high feedback systems show significantly lower
 central metallicities compared to the low feedback systems. If low
 entropy, metal-rich gas has been boosted to large entropy in the high
 feedback systems, it must now reside outside 0.5\,\RF, to remain
 undetected in our study. Considering gas as a function of scaled
 entropy, we find similar levels of enrichment in both high and low
 feedback systems, and argue that the lack of extra metals in the
 highest feedback systems points to an AGN origin for the bulk of the
 feedback, probably acting within precursor structures.
\end{abstract}


\section{Introduction}
\label{sec:intro}
The hierarchical view of structure formation indicates that present day
clusters were assembled from galaxy groups. Indeed, the majority of
galaxies in the Universe are situated in groups \citep{tully87}, so
understanding the processes at work in these systems is crucial to our
understanding of galaxy properties as a whole. Interestingly, groups also
seem to show a very diverse range of properties \citep[e.g.][]{osmond04},
and understanding the root causes of these differences is key to
understanding the processes at work in shaping present day galaxy
clusters. It has also been known for some time that groups do not obey
self-similarity, and for example, show a steepening in the $L_{x}-T_{x}$
relation compared to clusters \citep{helsdon00}. This is explained via the
increased importance of baryonic physics; cooling, feedback from AGN and
supernovae, merger shocks and galaxy winds can have significant effects 
on the properties of the gas within the shallower potential wells of low
mass clusters.

Knowledge of the dark matter-dominated potential, coupled with the entropy
distribution of the gas within a system in hydrostatic equilibrium,
completely defines the thermodynamic properties
of the intracluster medium (ICM) \citep{voit03b}. The advent of high
quality X-ray data from \Chandra\ and \XMM\ have allowed detailed studies
of the entropy properties of clusters \citep[e.g.][]{pratt06,morandi07} and
simulations have been extensively used to predict properties
such as entropy profiles and the chemical enrichment of the ICM
(see \citeauthor{borgani08a} 2008a,b \nocite{borgani08b} for recent reviews). However, the important
group regime has been less well explored, although significant progress is now 
being made in modelling feedback effects on the group scale \citep{dave08}. 
There is a parallel shortage of detailed {\it observational} evidence which 
bears on the processes shaping group evolution, and the aim of the present
study is in part to address this.

Observations of galaxy clusters show a striking bimodality in their
observed ICM properties, which leads to their classification into cool core (CC) and non cool core (NCC) systems. These classes are characterised by the
observation (or not) of central positive temperature gradients. The open 
question in this field is how this apparent bimodality is produced. The short
central cooling times observed in both types of system \citep{sanderson06},
causes difficulties for attempts to explain the two classes within the same
framework of cluster evolution. However, recent attempts have been made to
do just this \citep{mccarthy08}, by invoking different levels of
pre-heating before cluster collapse. 

Recent results show strong similarities between
CC clusters and CC groups. For example, the observed
abundance gradients in CC clusters \citep{degrandi04,baldi07} 
are also seen
in the \Chandra\ sample of 15 groups of \citet{rasmussen07}, 14 of which
were observed to have CCs. However, one crucial area for
understanding the role of feedback that has not yet been probed is the
nature of NCC systems in the group regime.

We are now entering an era where large studies of galaxy groups are
possible with high quality X-ray data
\citep{finoguenov06,finoguenov07,rasmussen07,sun08}. Although the sample
analysed here was not selected in a statistical manner, we have an
opportunity to gain considerable insight into the diversity of group
properties, as this is the largest sample of groups to date with high
quality \XMM\ data, which has been analysed in a homogeneous way. One key
advantage of our approach is that in the spectral analysis of this data,
spherical symmetry was not an \textit{a priori} requirement
\citep[see][]{finoguenov06,finoguenov07}. Additionally, due to the size of
the sample, we have been able, for the first time, to separate the sample
into CC and NCC groups, and also to look at the behaviour of groups as a
function of temperature. This provides a unique opportunity to examine
which processes drive the observed differences in group properties, by
considering the behaviour which diverges from the mean relations.

The layout of the paper is as follows. In Section \ref{sec:sample} we
describe the sample, analysis and basic properties of the groups, in
Section \ref{sec:divide} we divide the sample to look at the mean properties
of sub-samples of the data, and Section \ref{sec:radial} explores the
radial properties of the sample. We explore the entropy distributions and
discuss the effects of feedback on the groups in Section \ref{sec:Sdist},
and we summarise our main results in Section \ref{sec:conclude}.
\section{Group Sample}
\label{sec:sample}
The Two-Dimensional \XMM\ Group Survey (2dXGS) is an archival study of nearby (z $<$ 0.024) galaxy groups, which were selected from the group catalogue of \citet{mulchaey03} and were chosen to have publicly available \XMM\ data \citep{finoguenov06,finoguenov07}. The complete sample selection and data analysis is described by \citet{finoguenov06,finoguenov07}, hereafter referred to as F+06 and F+07 respectively. The former concentrates on the `low' redshift sample (z $<$ 0.012), and the latter concentrates on the `high' redshift sample (0.012 $<$ z $<$ 0.024). These works studied the radial properties of individual groups and their deviations from the mean profiles, applying a novel spectral approach which we briefly describe below. Our aim is to bring these two samples together, supplemented by groups from the sample of \citet{mahdavi05} which were analysed using the same procedures, in an effort to undertake a full statistical study of the properties of a large sample of galaxy groups derived from high quality X-ray observations. 

The full 2dXGS sample contains 25 nearby galaxy groups observed with \XMM,
which have been analysed in a homogeneous way. A summary of the data
analysis procedure is given here; we refer the reader to F+06 and F+07, and
references therein, for a full description of the \XMM\ data, reduction and
analysis. There are two main stages in the analysis following the initial
\XMM\ data reduction. Firstly, temperature and surface brightness maps were
used to look at the overall structure of the group. The second part of the
analysis extracted spectra from regions of contiguous surface brightness
and temperature, and fitted single temperature hot plasma (\textsc{apec})
models to the spectra, yielding the spectral properties of the group with
no \textit{a priori} assumption of the system being spherically
symmetric. The abundances used were those of \citet{and89}, and absorption
was fixed at the Galactic value.

The result of this approach is to yield a series of spectrally derived
parameters in both two-dimensional regions, and also from a more
traditional spectral analysis of a series of concentric annuli. We can look
at the radial properties of each group by assuming the characteristic
radius for the measurement to be the mean radius of the region. Here we
will concentrate mainly on using the properties derived from this novel
analysis of 2d regions, except where otherwise stated. Due to the nature of
the analysis, the customary spherical deprojection of the spectra cannot be
performed. Instead, F+06 and F+07 determined the three dimensional gas
properties in each of the analysed regions, by estimating the projection
length of each region. We use here the derived gas properties from this
approach, and refer readers to F+06, F+07 and references therein for more
details on this procedure.

We define the mean temperature of each group to be that recovered from
fitting a spectrum extracted from the radial range 0.1--0.3\,\RF\ (F+06,
F+07), so we restrict the analysis presented here to the 21 2dXGS groups
with measured temperatures in this range. We have supplemented the 21
groups in the 2dXGS sample with seven groups with the highest data quality
from \citet{mahdavi05}, leading to a final sample of 28 galaxy groups. The
\citet{mahdavi05} groups were analysed using the same two-dimensional
procedure, but they have not also been analysed with the traditional annular
approach which was also applied in the
case of the 2dXGS groups. The groups in the
Mahdavi sample were initially selected from the \Rosat\ All-Sky
Survey/Center for Astrophysics Loose Systems (RASSCALS) of
\citet{mahdavi00}, and cover redshifts between 0.016 and 0.037. The mean
temperatures for these groups have been re-extracted, to cover the
0.1--0.3\,\RF\ region applied to the 2dXGS groups.

\subsection{Group properties}
The basic properties of the group sample appear in Tables \ref{table:main} and \ref{table:main2}, the first of which refers to the 2dXGS sample of F+06 and F+07, and the second of which shows the properties of the supplementary groups from \citet{mahdavi05}. Column (1) shows the group name as expressed by F+06, F+07 and \citet{mahdavi05}. We now describe the origin of the values shown in Tables \ref{table:main} and \ref{table:main2}. We assume a Hubble constant of \Hn\ = 70~kms$^{-1}$Mpc$^{-1}$ throughout. 

\subsubsection{2dXGS groups}
\label{sec:2dXGS}
Due to the overlap of 15 of the 2dXGS groups with the Group Evolution
Multiwavelength Study (GEMS) sample of \citet{osmond04}, the latter is our
primary source for the group properties (shown in Table~\ref{table:main}) 
supplementary to those provided by F+06 and F+07. Where distance
measurements were available from \citet{osmond04}, these were used. For the
remaining groups, we re-scaled the distance measurements of
\citet{mulchaey03} to our assumed Hubble constant (3C~499, NGC~507,
NGC~2300 and NGC~4168). In the case of Hickson 51 and the Pavo group, it
was necessary to estimate the distance from the redshift information
provided in F+07.

X-ray luminosities are also drawn primarily from \citet{osmond04}, which 
employed \Rosat\ PSPC data, and has the advantage of extending to
larger radii than \XMM\ data for many of these groups.
We used the supplied $\beta$-model
surface brightness fits of \citet{osmond04} to re-scale the X-ray
luminosities from their original extraction radius to our \RF\ values. 
Individual extraction radii can be found in \citet{osmond04}; they
are typically $\sim$\,55 percent of our \RF.
In some cases, Osmond \& Ponman could not fit a model to 
the surface
brightness profile, and in these cases, a standard model of $\beta$ = 0.5
and r$_{core}$ = 6~kpc was assumed. This affects the following groups in
our sample: Hickson 15, Hickson 92 and NGC~5171. For groups not in the GEMS
sample, X-ray luminosities are from \citet{mulchaey03}, but have been
corrected in the following way for differences in the assumed Hubble
constant, and to extrapolate to \RF\ in line with the X-ray luminosities of
the GEMS groups. Where $\beta$-model surface brightness fits were available
from \citet{mulchaey03} (NGC 507), we re-scaled the X-ray luminosities both
for the assumption here of a lower Hubble constant, and to scale from the
original extraction radius of 200~h$_{100}^{-1}$~kpc to our values of
\RF. In two cases (3C~449, NGC~2300), the fitted core radii reported by
\citet{mulchaey03} are lower than the resolution limit of the \Rosat\ PSPC
instrument. In this case we set the core radius at this limit when scaling
the X-ray luminosity to our value of \RF. For NGC~4168, where only an upper
limit on $L_{x}$ was determined by \citet{mulchaey03} and hence no surface
brightness modelling was undertaken, we report the upper limit corrected
for the different Hubble constant only. In the case of Hickson 51, which is
not in the GEMS sample, we used the X-ray luminosity for the group NRGb128
from the RASSCALS sample of \citet{mahdavi00}, which is given as an
equivalent designation for Hickson 51 in the NASA/IPAC Extragalactic
Database (NED)\footnote{http://nedwww.ipac.caltech.edu/}. We extrapolated
this X-ray luminosity from the original extraction radius of
0.5~h$_{100}^{-1}$\,Mpc to our \RF\ assuming the standard $\beta$-model of
\citet{osmond04}, and we also scaled the X-ray luminosities to a Hubble
constant of \Hn\ = 70~kms$^{-1}$Mpc$^{-1}$. This is the same procedure we
have used to determine X-ray luminosities for the majority of the
\citet{mahdavi05} groups (see Section \ref{sec:mahdavi}), so there is a
consistency in our approach.

It was necessary to estimate the X-ray luminosity of Pavo from the work of
\citet{machacek05}, who fitted a $\beta$-model surface brightness profile
to the extended emission outside the central galaxy (NGC 6876) with
r$_{core}$ = 196$^{\prime\prime}$ and $\beta$ = 0.3, yielding an X-ray
luminosity (in the energy band 0.5--2~keV) of log $L_{x}$ = 41.8 within
463$^{\prime\prime}$, assuming a Hubble constant of
75~kms$^{-1}$Mpc$^{-1}$. We have extrapolated this model to our \RF\ and
have corrected for the differences in the assumed Hubble constant, to quote
the X-ray luminosity for Pavo shown in Table \ref{table:main}. The
remaining caveat with this value is the slightly different energy band
compared to the remainder of the sample, which were all derived from \Rosat\
data.

The velocity dispersion of each group comes from F+06 and F+07, except in
the case of NGC 4636, for which F+06 could not determine the group
membership satisfactorily (we refer the reader to this work for more
information), and in this case, we quote the velocity dispersion of
\citet{osmond04}. The mean temperatures of the groups and the values of
\RF\ were all derived by F+06 and F+07 using the method described in
Section \ref{sec:sample}.

We also show selected optical properties of the central galaxy and the
group. Values of \D, the diameter of the isophote where the $B$-band
surface brightness is 25~mag~arcsec$^{-2}$, are determined for the central
group galaxy from the RC3 catalogue of \citet{RC3}. We have applied the
procedure of \citet{osmond04} in determining the $B$-band luminosity of the
brightest group galaxy (BGG) and of the group, by extracting galaxies from
NED within a projected radius of \RF, centred on the group
co-ordinates. For groups in the \citet{osmond04} sample, we used the group
co-ordinates provided in this work, and for the remaining groups we use the
NED co-ordinates corresponding to the group name. The group luminosities
are 90 percent complete and we applied an absolute magnitude cut of $M_{B}$
= -16.32 to the group galaxies. The BGG is chosen to be the brightest galaxy 
within 0.25\,\RF\ of the group centre. We refer the reader to
\citet{osmond04} for more information on the applied method. In the case of
the Pavo group, we have assumed the BGG to be NGC 6876 \citep{machacek05};
this is confirmed by our procedure for extracting $B$-band
luminosities. Table~\ref{table:main} also gives the maximum radius 
in units of \RF\ to which spectral information is available, for each of the 
groups, to indicate the completeness of the spectral coverage.
The final columns in Table~\ref{table:main}
denote the subsamples to which each group
belongs, in terms of their mean temperatures and core properties. These
classifications are defined in Section \ref{sec:divide}.
\begin{table*}
  \centering
  \caption{The basic properties of the 2dXGS groups in the sample. The groups have been classified by their mean temperature and their core properties (see Section \ref{sec:divide}). Distances and values of \LX\ are from \citet{osmond04} unless otherwise stated. For data from \citet{osmond04}, the X-ray luminosities have been extrapolated to our \RF\ values (see text). \ensuremath{L_{B,BGG}} and \ensuremath{L_{B,grp}} are calculated as described in the text. Mean temperatures and values of \RF\ are from F+06 and F+07. Velocity dispersions are also from F+06, F+07, except in the case of NGC 4636, where the value used is that from \citet{osmond04}, and values of \D\ (the diameter of the isophote where the surface brightness is 25~mag/arcsec$^{2}$ in the $B$-band) are from RC3 \citep{RC3}, and refer to the brightest group galaxy (BGG). $r_{max}$ denotes the maximum radius to which spectral information is available, in units of \RF.}
  \label{table:main}
  \begin{tabular}{ccccccccccccc}
  \hline
     Group & Dist. & log \ensuremath{L_{x}} & $\sigma$ & \ensuremath{\bar{T}} & \ensuremath{r_{500}} & \D\ & log \ensuremath{L_{B}} & log \ensuremath{L_{B}} & $r_{max}$ & Warm/Cool & CC/NCC\\
       & (Mpc) & (ergs$^{-1}$) & (kms$^{-1}$) & (keV) & (kpc) & (arcmins) & (L$_{B,\odot}$) & (L$_{B,\odot}$) & (\RF) &\\
   & & & & & & & BGG & Group & & &\\
  \hline
  3C 449 & 67$^{a}$ & 42.93$\pm^{0.26}_{0.27}$$^{a}$ & 335$\pm112$ & 1.28$\pm$0.02 & 453 & 1.10 & 10.30 & 10.77 & 0.45 & Warm & NCC\\
  HCG 15 & 95 & 42.19$\pm$0.05 & 404$\pm$122 & 0.62$\pm$0.04 & 286 & 0.87 & 10.04 & 10.66 & 0.52 & Cool &  NCC\\
  HCG 42 & 64 & 42.05$\pm$0.02 & 282$\pm$43 & 0.75$\pm$0.19 & 324 & 2.95 & 10.95 & 11.27 & 0.28 & Cool & CC\\
  HCG 51 & 114$^{b}$ & 42.65$\pm$0.13$^{c}$ & 546$\pm$151 & 1.16$\pm$0.13 & 424 & 1.10 & 10.51 & 11.10 & 0.82 & Warm & NCC\\
  HCG 62 & 74 & 43.18$\pm$0.04 & 418$\pm$51 & 1.06$\pm$0.02 & 403 & --$^{d}$ & 10.54 & 11.32 & 0.46 & Warm & CC\\
  HCG 68 & 41 & 41.71$\pm$0.04 & 191$\pm$68 & 0.69$\pm$0.09 & 308 & 2.19 & 10.63 & 11.32 & 0.19 & Cool & NCC\\
  HCG 92 & 88 & 42.16$\pm$0.04 & 467$\pm$176 & 0.79$\pm$0.24 & 334 & 1.91 & 10.51 & 11.01 & 0.42 & Cool & NCC\\
  IC 1459 & 26 & 41.42$\pm$0.04 & 256$\pm^{44}_{38}$ & 0.59$\pm$0.03 & 280 & 5.25 & 10.63 & 10.90 & 0.21 & Cool & NCC\\
  NGC 507 & 70$^{a}$ & 43.37$\pm$0.03$^{a}$ & 580$\pm$94 & 1.34$\pm$0.01 & 467 & 3.09 & 11.00 & 11.60 & 0.53 & Warm & CC\\
  NGC 533 & 76 & 42.71$\pm$0.03 & 439$\pm$60 & 1.26$\pm$0.01 & 448 & 3.80 & 10.99 & 11.39 & 0.56 & Warm & CC\\
  NGC 2300 & 29$^{a}$ & 41.93$\pm$0.04$^{a}$ & 278$\pm^{35}_{31}$ & 0.75$\pm$0.01 & 339 & 2.81 & 10.34 & 10.67 & 0.32 & Cool & CC\\
  NGC 2563 & 73 & 42.58$\pm$0.03 & 384$\pm$49 & 1.31$\pm$0.05 & 460 & 2.09  & 10.64 & 11.35 & 0.36 & Warm & CC\\
  NGC 4073 & 96 & 43.46$\pm$0.02 & 565$\pm$72 & 1.87$\pm$0.05 & 575 & 3.16 & 10.96 & 11.65 & 0.49 & Warm& CC\\
  NGC 4168 & 38$^{a}$ & $<$40.87$^{a}$ & 259$\pm^{56}_{52}$ & 0.77$\pm$0.31 & 258 & 2.75 & 10.39 & 10.81 & 0.19 & Cool & NCC\\
  NGC 4261 & 41 & 42.33$\pm$0.03 & 429$\pm^{54}_{50}$ & 1.11$\pm$0.02 & 451 & 4.07 & 10.85 & 11.46 & 0.19 & Warm  & CC\\
  NGC 4325 & 117 & 43.16$\pm$0.01 & 376$\pm$70 & 1.01$\pm$0.01 & 389 & 0.95 & 10.65 & 10.97 & 0.54 & Cool & CC\\
  NGC 4636 & 10 & 41.72$\pm$0.02 & 284$\pm$73 & 0.77$\pm$0.01 & 331 & 6.03 & 10.02 & 10.39 & 0.18 & Cool & CC\\
  NGC 5044 & 33 & 43.10$\pm$0.01 & 357$\pm^{48}_{42}$ & 1.21$\pm$0.01 & 430 & 2.95 & 10.50 & 11.15 & 0.40 & Warm & CC\\
  NGC 5171 & 107 & 42.48$\pm$0.06 & 494$\pm$99 & 1.21$\pm$0.05 & 436 & 1.10 & 10.77 & 11.44 & 0.49 & Warm & NCC\\
  NGC 5846 & 30 & 42.01$\pm$0.02 & 368$\pm^{51}_{46}$ & 0.69$\pm$0.01 & 309 & 4.07 & 10.73 & 11.10 & 0.13 & Cool & CC\\
  Pavo & 57$^{b}$ & 42.69$^{e}$ & 440$\pm$96 & 0.77$\pm$0.12 & 330 & 2.82$^{f}$ & 11.02 & 11.47 & 0.49 & Cool & NCC\\ 
  \hline
  \end{tabular}
 \begin{list}{}{}
  \item[Notes:]
  \item[$^{a}$] Values from \citet{mulchaey03}, rescaled using \Hn~=~70~kms$^{-1}$~Mpc$^{-1}$ and extrapolated to our \RF\ values (see text for details). 
  \item[$^{b}$] Distance estimated from redshift given in \citet{finoguenov07}.
  \item[$^{c}$] Value from \citet{mahdavi00} rescaled using \Hn~=~70~kms$^{-1}$~Mpc$^{-1}$ and extrapolated to our \RF\ values using the standard $\beta$-model of \citet{osmond04} (see text for details). 
  \item[$^{d}$] Not available in RC3.
  \item[$^{e}$] $L_{x}$ extrapolated to \RF\ and corrected for \Hn~=~70~kms$^{-1}$~Mpc$^{-1}$ from the measured X-ray luminosity and surface brightness profile of \citet{machacek05} (Note: energy band is 0.5--2~keV for this value).
  \item[$^{f}$] \D\ from RC3 assuming central galaxy is NGC 6876 \citep{machacek05}. 
  \end{list}
\end{table*}
\subsubsection{\citet{mahdavi05} groups}
\label{sec:mahdavi}
Three of the \citet{mahdavi05} groups appear in the sample of \citet{osmond04}. These are Hickson 97, SRGb119 \citep[central galaxy NGC 741;][]{mahdavi05} and NGC 5129. For these groups, distances are from \citet{osmond04}, but in the remainder of cases the distances were estimated from the redshifts presented by \citet{mahdavi05}. Similarly, the X-ray luminosities for these 3 groups are re-scaled from the original extraction regions presented in \citet{osmond04} to match our \RF\ values, as described in Section \ref{sec:2dXGS} for the groups in the 2dXGS sample. For the remaining four groups, we use the X-ray luminosities from \citet{mahdavi00}, which were extracted from the \Rosat\ All-Sky Survey, within a radius of 0.5~h$_{100}^{-1}$Mpc. Assuming the standard $\beta$-model of \citet{osmond04}, we scale these X-ray luminosities both for \Hn\ = 70~kms$^{-1}$Mpc$^{-1}$ and to our radius of \RF, in the manner described in Section \ref{sec:2dXGS}. This ensures a degree of consistency between the 2dXGS and \citet{mahdavi05} samples. 

Velocity dispersions for all seven systems come from \citet{mahdavi05}, and mean temperatures and \RF\ values were re-derived from \citet{mahdavi05} in the radial range 0.1--0.3\,\RF. The $B$-band optical luminosities of the BGG ($L_{B,BGG}$) and the group ($L_{B,group}$) were determined using the same procedure as in Section \ref{sec:2dXGS}. For the three GEMS groups, we used the co-ordinates provided by \citet{osmond04} for the NED search; we use the group co-ordinates from \citet{mahdavi05} for the remainder of the sample. \citet{mahdavi05} indicate that Abell 194 has no dominant galaxy, but our method finds NGC 541 to be the brightest galaxy within 0.25\,\RF, and we adopt this as the BGG. The central galaxy of RGH~80 \citep[NGC 5098;][]{mahdavi05} is listed in NED as a galaxy pair, but we find PGC 046515 to be the brightest galaxy within 0.25\,\RF\ of the group centre, and we adopt this galaxy as the BGG in this system. Values of \D\ were again extracted from the RC3 catalogue of \citet{RC3} for these systems. Table~\ref{table:main2} also gives the maximum radius 
in units of \RF\ to which spectral information is available, for each of the 
groups, to indicate the completeness of the spectral coverage. \\

\begin{table*}
  \centering
  \caption{The basic properties of the groups in the sample originally from \citet{mahdavi05}. Mean temperatures and values of \RF\ have been re-derived from \citet{mahdavi05} in the radial range 0.1--0.3\,\RF. Distances have been estimated from the redshifts given in \citet{mahdavi05}, and X-ray luminosities have been extrapolated to our \RF\ values from \citet{osmond04} and \citet{mahdavi00}. Values of \ensuremath{L_{B,BGG}} and \ensuremath{L_{B,grp}} are calculated as described in the text. \D\ values are from RC3 \citep{RC3}. $r_{max}$ denotes the maximum radius to which spectral information is available, in units of \RF.}
  \label{table:main2}
  \begin{tabular}{cccccccccccc}
  \hline
     Group & Dist. & log \ensuremath{L_{x}} & $\sigma$ & \ensuremath{\bar{T}} & \ensuremath{r_{500}} & \D\ & log \ensuremath{L_{B}} & log \ensuremath{L_{B}} & $r_{max}$ & Cool/Warm & CC/NCC\\
       & (Mpc) & (ergs$^{-1}$) & (kms$^{-1}$) & (keV) & (kpc) & (arcmins) & (L$_{B,\odot}$) & (L$_{B,\odot}$) & (\RF) & &\\
   & & & & & & & BGG & Group & & &\\
  \hline
  A 194 & 76$^{a}$ & 42.88$\pm$0.05$^{b}$ & 550$\pm^{89}_{86}$ & 1.01$\pm$0.15 & 393 & 1.78$^{c}$  & 10.74$^{c}$  & 11.56 & 0.86 & Cool & NCC\\
  HCG 97 &  92 & 42.77$\pm$0.10 & 383$\pm^{50}_{52}$ & 1.20$\pm$0.05 & 439 & 1.66 & 10.39 & 11.05 & 0.69 & Warm & CC\\
  NGC 5129 & 108 & 43.14$\pm$0.09 & 283$\pm^{29}_{31}$ & 0.95$\pm$0.03  & 379 & 1.70 & 11.05 & 11.37 & 0.53 & Cool & CC\\
  NRGb 184 & 96$^{a}$ & 42.57$\pm$0.12$^{b}$ & 390$\pm^{40}_{37}$ & 1.37$\pm$0.09 & 477 & 0.87 & 10.41 & 11.04 & 0.68 & Warm & CC\\
  RGH 80 & 158$^{a}$ & 43.18$\pm$0.06$^{b}$ & 602$\pm^{63}_{61}$ & 1.16$\pm$0.02 & 429 & 0.78$^{d}$ & 10.59$^{d}$ & 11.37 & 1.4 & Warm & CC\\
  SRGb 119 & 79 & 42.65$\pm$0.10 & 416$\pm^{34}_{33}$ & 1.34$\pm$0.07 & 470 & 2.95 & 11.11 & 11.29 & 0.58 & Warm & CC\\
  SS2b153 & 68$^{a}$ & 42.69$\pm$0.07$^{b}$ & 161$\pm^{19}_{71}$ & 0.83$\pm$0.01 & 348 & 2.09 & 10.61 & 10.89 & 0.68 & Cool & CC\\
\hline
\end{tabular}
\begin{list}{}{}
  \item[Notes:]
  \item[$^{a}$] Distances estimated from redshifts given in \citet{mahdavi05}.
  \item[$^{b}$] Values from \citet{mahdavi00} extracted within 0.5~Mpc, rescaled here for \Hn~=~70~kms$^{-1}$~Mpc$^{-1}$ and extrapolated to our \RF\ using the standard $\beta$-model of \citet{osmond04} (see text for details).
  \item[$^{c}$] NGC 541 assumed as BGG (see text for details) as no dominant galaxy listed by \citet{mahdavi05}.
  \item[$^{d}$] PGC 046515 assumed as BGG (see text for details).
\end{list}
\end{table*}
\section{Dividing the Sample}
\label{sec:divide}
The 2dXGS sample of groups has been presented previously (F+06, F+07)
in two subsamples, distinguished by group distance,
but it is also instructive to partition the sample
on the basis of {\it physical} properties. We have divided the
groups into two main subsamples, firstly on the basis of the mean
temperature of the group, which acts as a proxy for the group mass, and
secondly on the basis of whether or not the group has a CC. The
classification of galaxy clusters into CC and NCC
systems is well-known \citep[e.g.][]{peres98}. These systems have
been shown to exhibit different observational properties
\citep[e.g.][]{sanderson06}, and hence it is instructive to extend this
classification to the group regime. \citet{rasmussen07} studied the
temperature and abundance profiles of 15 groups with \Chandra\ data, 14 of
which were noted to have a CC. The properties of groups that do not
have CCs, and the connection between the overall state of the ICM
and the core properties of groups, has not been investigated to date. A
sample of the size presented here, although not statistically selected,
offers an opportunity to probe the wider properties of groups compared to
the presence (or not) of a CC.

To enable us to compare the radial profiles of groups directly, we scale 
the radii by \RF, the radius within which the mean density of the group 
is 500 times the critical density. This was calculated in an 
iterative fashion using (F+06, F+07), 
\begin{equation}
\label{eqn:r500}
r_{500} = 0.391 \bar{T}~^{0.63}h_{70}^{-1}.
\end{equation}
The partition of the groups into subsamples based on
their mean temperature and core properties is detailed below.

\subsection{Temperature}
\label{sec:temp_class}
We define the mean temperature of each group as that determined within the
radial range 0.1--0.3\,\RF\ (F+06, F+07). This removes the effect of any
CC on the temperature determination, providing a robust measure of
the mean temperature of the system. The median temperature across all
groups was found to be 1.035~keV, and this was used to divide the groups
into 2 subsamples, the \textit{Cool} groups with $\bar{T}$ $<$ 1.035~keV,
and the \textit{Warm} groups with $\bar{T}$ $\ge$ 1.035~keV. The
temperature classification for each group appears in Tables
\ref{table:main} and \ref{table:main2}.

\subsection{Core properties}
\label{sec:core_class}
\begin{figure*}
\includegraphics[width=15cm]{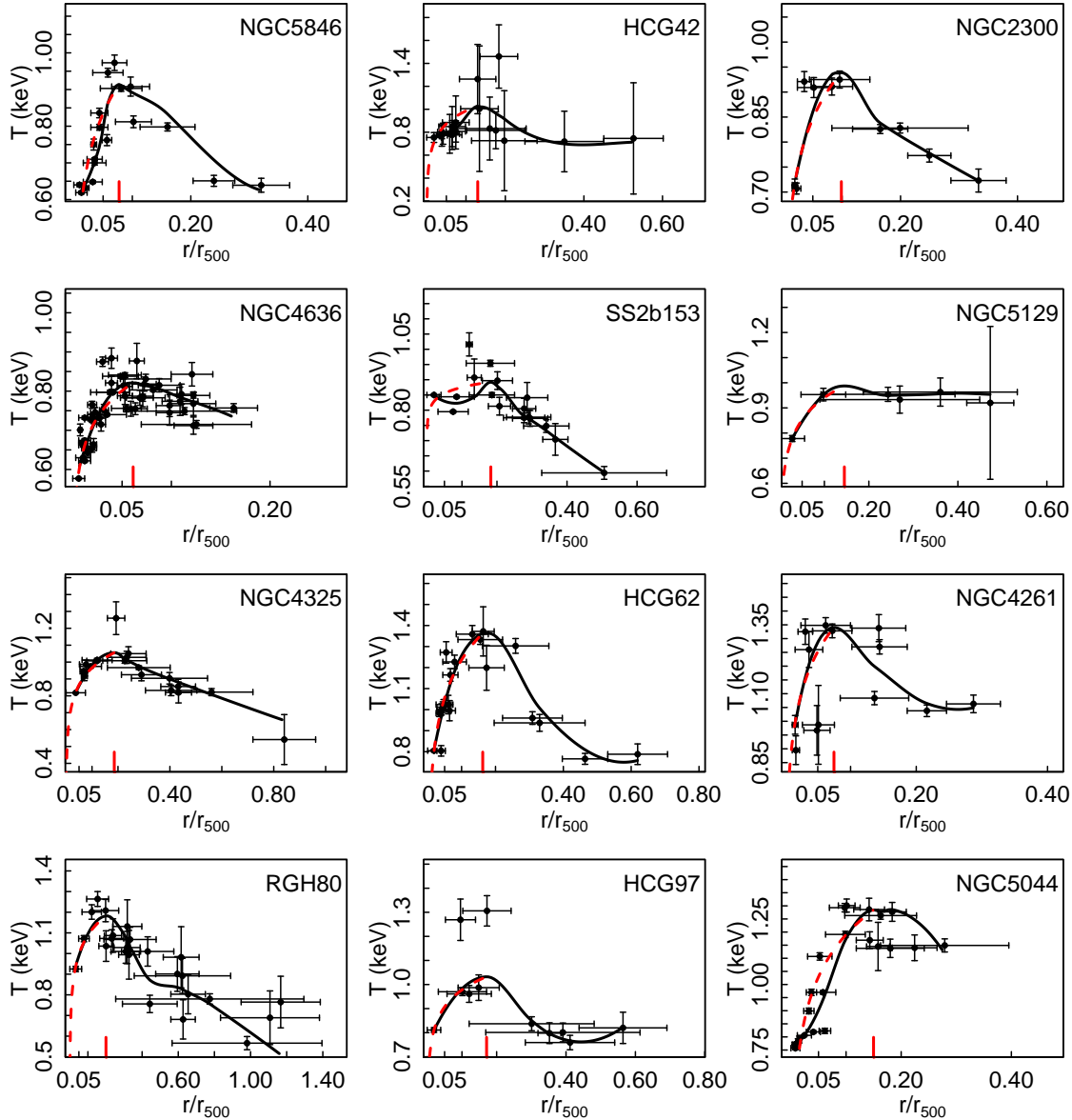}
\caption{The temperature profiles of the groups classified as being CCs. In the case of the 2dXGS groups, the data from both the annuli and two-dimensional regions has been used. The solid lines show weighted local regression fits to the data, see the text for details. The plots are arranged in order of ascending mean temperature, from top left to bottom right. Red tick marks show the radius of maximum temperature used in the calculation of the temperature drop, and the dashed red line shows a straight line fit (fitted in log-log space) to the temperature profiles in the core region. \textit{Continued overleaf.}}
\label{fig:kTCC}
\end{figure*}
\setcounter{figure}{0}
\begin{figure*}
\includegraphics[width=15cm]{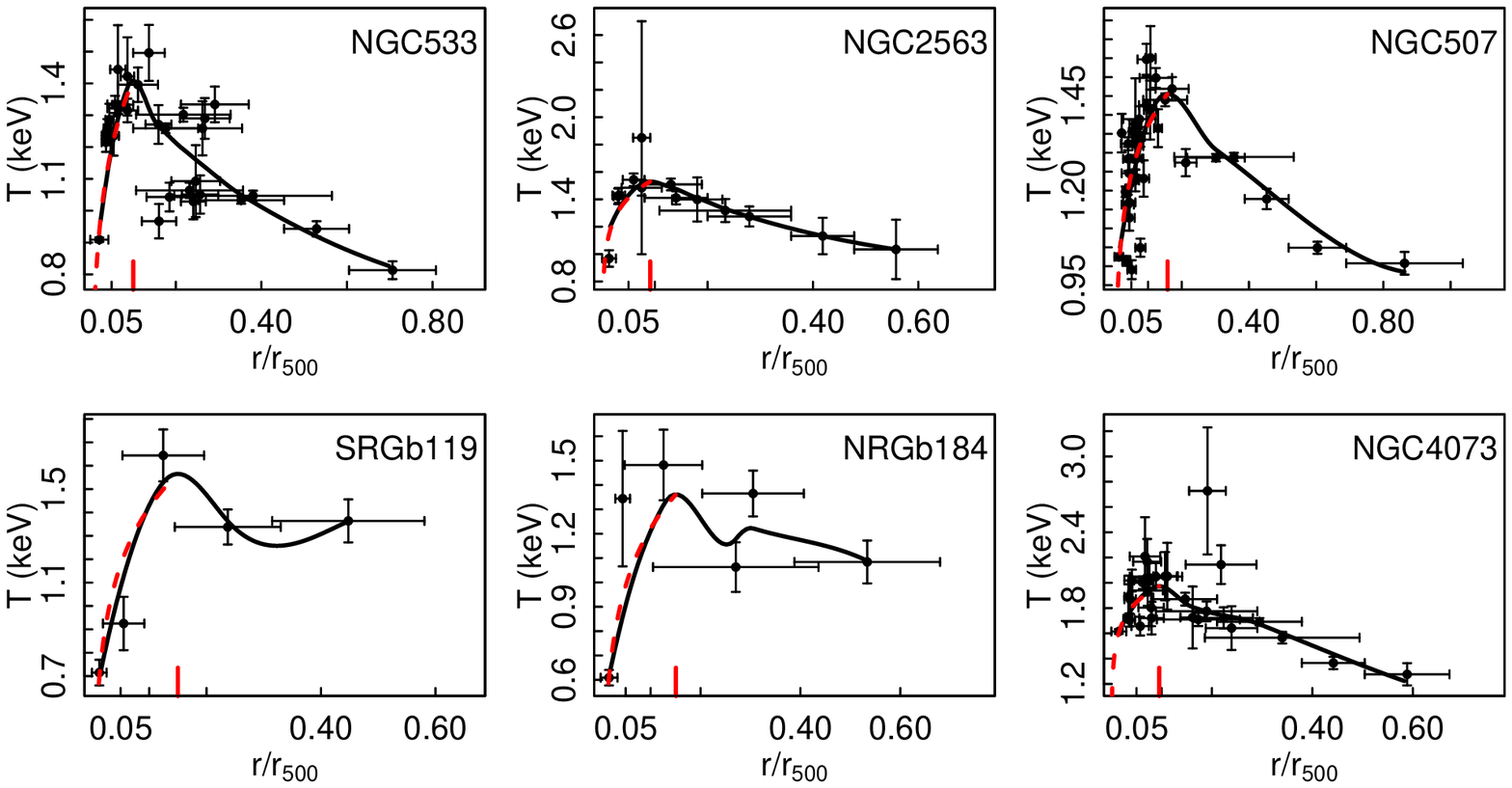}
\caption{\textit{Continued from previous page.}}
\end{figure*}
\begin{figure*}
\includegraphics[width=15cm]{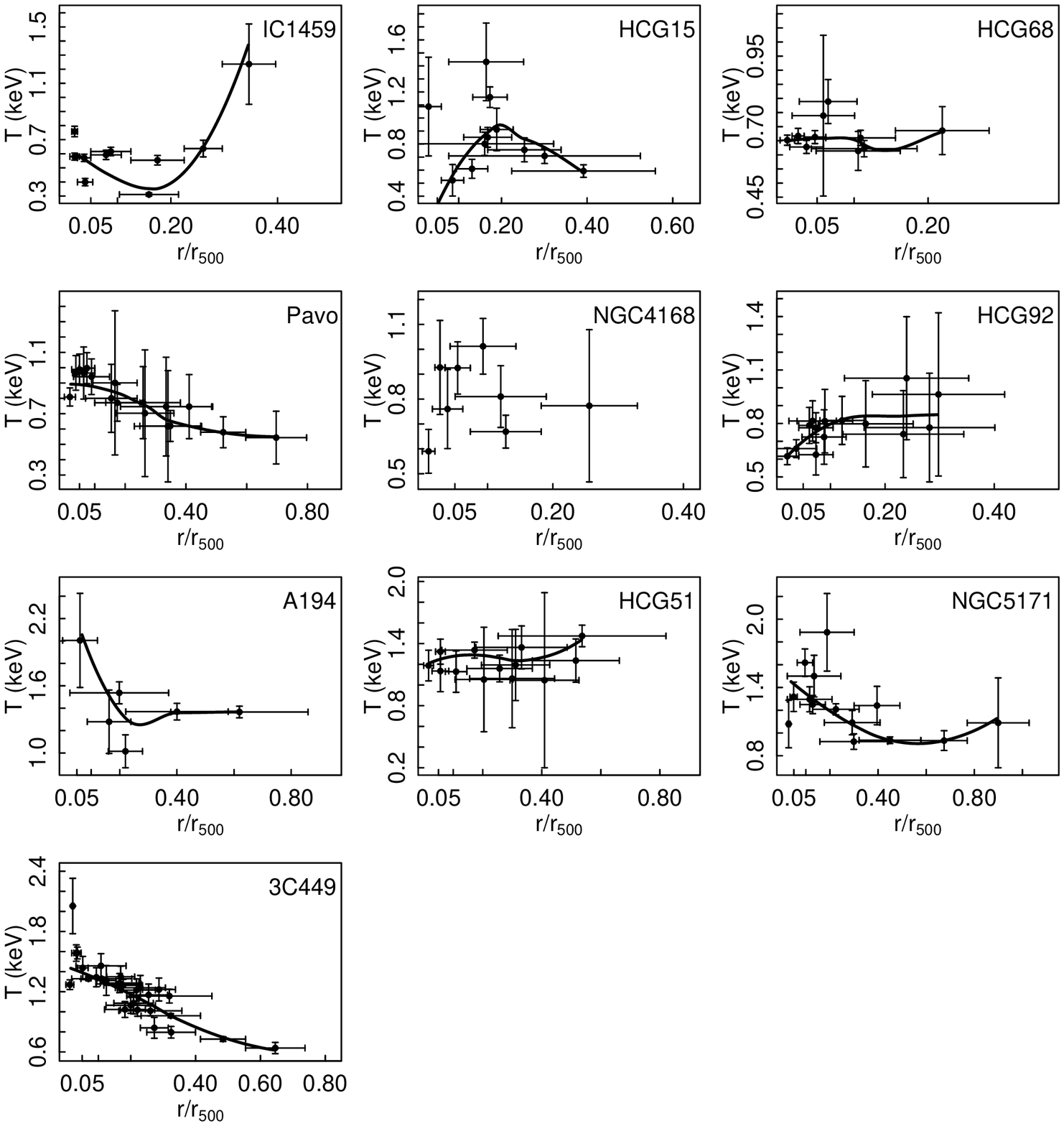}
\caption{The temperature profiles of the groups classified as being NCCs. In the case of the 2dXGS groups, the data from both the annuli and two-dimensional regions have been used. The plots are arranged in order of ascending mean temperature, from top left to bottom right. The solid lines show weighted local regression fits to the data, see the text for details. We note here that no smooth fit could be found for the temperature profile of NGC 4168, hence a fit is not shown in this case. }
\label{fig:kTNCC}
\end{figure*}
To distinguish between CC and NCC systems, one traditional approach is to
consider the cooling time of the system \citep[e.g.][]{peres98}. However,
\citet{sanderson06} showed that in the cluster regime, systems with no central 
cooling can also exhibit the short 
central cooling times ($<$~5\,Gyr) typically associated with CC
systems. They defined the presence of a CC by considering the ratio of
temperatures between an inner radius and an intermediate radial range, and
we adopt the same approach here. 

We define a group to have a CC if the
mean temperature in the region 0.1--0.3\,\RF\ is greater than the mean
temperature in the range 0.00--0.05\,\RF, indicating a positive temperature
gradient in the inner regions. If this was not the case, the group is
classified as a NCC system. This classification is based wherever possible
on results
from the analysis of both the annular and the 2d regions reported in the
earlier 2dXGS papers, to improve the radial coverage and allow a more
thorough appraisal of the temperature profile. For the seven groups from 
the \citet{mahdavi05} sample, only the 2d information is available. The
temperature profiles of the CC systems are shown in Figure \ref{fig:kTCC},
and those for the NCC systems are shown in Figure \ref{fig:kTNCC}. In both
cases, a weighted local regression fit to the data is also shown, to
indicate the overall behaviour of the temperature profile whilst
suppressing scatter. This fit was performed using the `\textsc{loess}'
function in version 2.5.1 of the \rr\ statistical environment
package\footnote{http://www.r-project.org} \citep{Rcite}, hereafter
referred to as \rr. The \textsc{loess} algorithm performs a weighted least
squares fit in the local neighbourhood of each data point, where the size of
the neighbourhood is defined to include a specified proportion of the data
sample, and the distance to each neighbour is used to weight the least 
squares fit. For more information on the \textsc{loess} algorithm, we 
refer the reader to \citet{cleveland92}.

For two systems, classification on the basis of the above criterion 
was problematical. Hickson 92 would be designated a CC system, since
its profile shows a small drop in central temperature, albeit of rather
low significance. However, this group (commonly known as Stephan's 
Quintet) is a system of galaxies undergoing multiple collisions, which
are currently disturbing and heating its intergalactic
medium \citep{trinchieri03,trinchieri05}. It is far from being a 
typical CC system, in which radiatively cooling, dense gas
is centred on a dominant early-type galaxy. We therefore reclassify
it as an NCC system. In the case of Abell 194, the quality and 
resolution of the data did
not permit us to extract a central spectrum within 0.05\,\RF.
However, the temperature profile rises consistently inward
to the innermost point,
at $\sim$0.06\,\RF. We are therefore confident in classifying
this group as a NCC, since all our CC systems
show CC behaviour by this radius,
whereas Abell 194 does not. The final sample therefore consists of 10 NCC
groups, and 18 CC groups. Ten of the latter are included in the 
\Chandra\ study of \citet{rasmussen07}.

One interesting property of CC groups is the magnitude of the drop in
temperature within the core. To estimate this, we found the radius of
peak temperature, from the maximum in the local
regression fits to the data. To be able to evaluate the temperature drop to
the same small radius (0.01\,\RF) in all groups, we fitted a straight line
in log-log space to the temperature within the radius of maximum
temperature (red dashed lines in Figure \ref{fig:kTCC}), forcing the
condition of passing through this maximum point. This fit was then used
to evaluate the temperature
at 0.01\,\RF, allowing groups with different radial sampling
to be directly compared. 

For our CC subsample, the
ratio of the temperature at 0.01\,\RF\ to the mean temperature of the 
system is found to be $T_c/\bar{T}$ = 0.69$\pm$0.16. This
agrees with the result (0.58$\pm$0.14) of
\citet{rasmussen07} within errors. However, the `central temperatures'
of \citep{rasmussen07} were measured in the innermost radial bin of their
analysis, which in most cases lay within 0.01\RF. This may account
for their slightly lower value of $T_c/\bar{T}$.

\begin{figure}
\includegraphics[width=8cm]{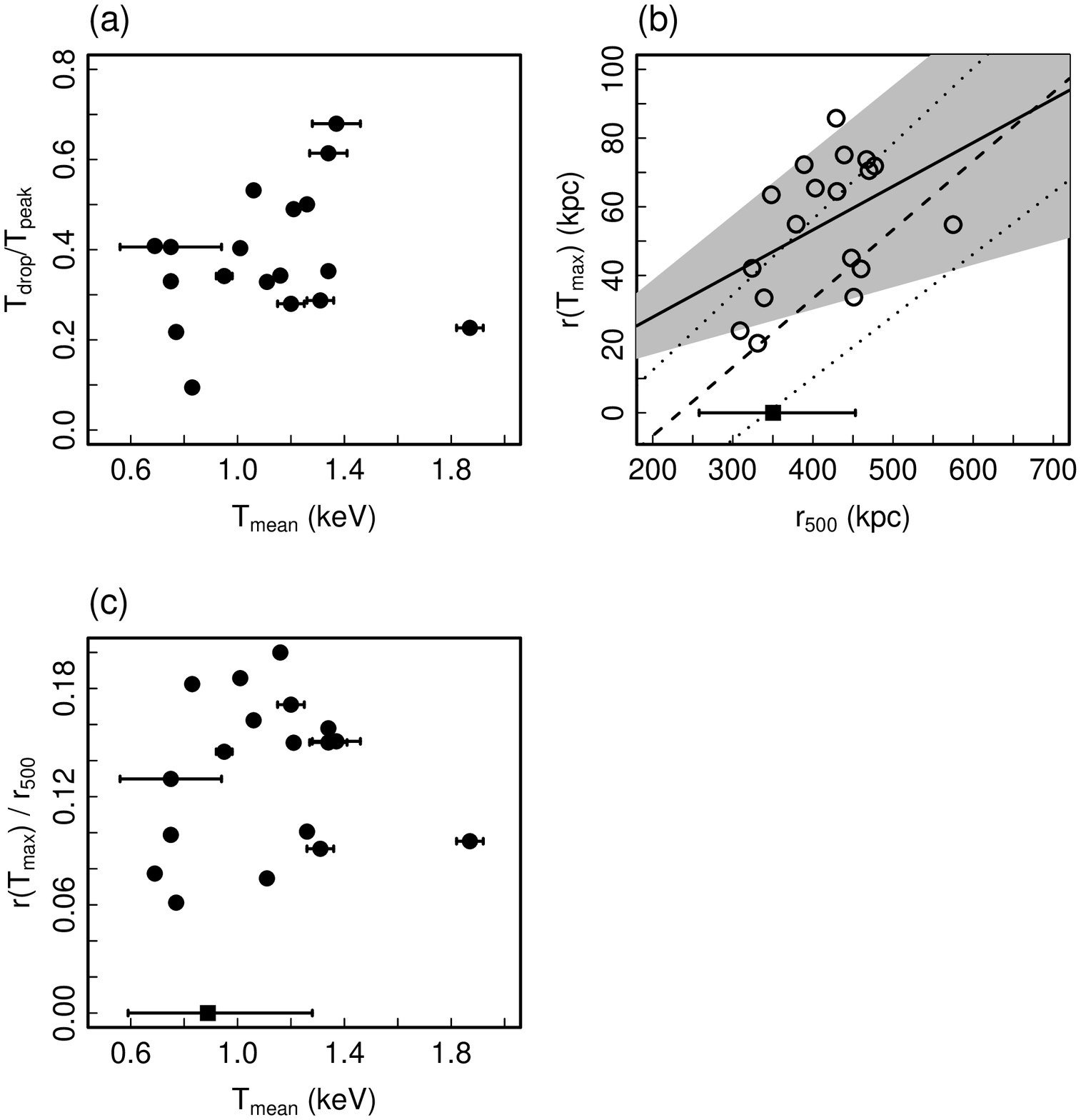}
\caption{\textit{(a):} The fractional temperature drop (T$_{drop}$\,/\,T$_{peak}$) in the CC systems versus the mean temperature of the system. \textit{(b):} The radius of the maximum temperature versus \RF\ for the CC groups (open circles). The dashed line is the fit of \citet{rasmussen07} to their sample of CC groups (excluding NGC 4125) analysed using \Chandra\ data, with the dotted lines showing the upper and lower bounds of the fit. The solid line is an orthogonal regression fit to our data (see text), with its 1\,$\sigma$ associated confidence region. The solid square marks the mean of the \RF\ values for the NCC groups, and the associated error bar marks the minimum and maximum of the \RF\ values for the NCC groups in the sample. \textit{(c):} The radius of maximum temperature (as a fraction of \RF) versus the mean temperature of the system. The solid square marks the mean of the \RF\ values for the NCC groups, and the associated error bar marks the minimum and maximum of the \RF\ values for the NCC groups in the sample.}
\label{fig:rmax}
\end{figure}

We quantified the observed temperature drop as the difference between the
maximum temperature and that interpolated at 0.01\,\RF. This quantity is
temperature dependent, as hotter systems have the capacity to show larger
temperature drops. To remove this dependence, we divided the temperature
drop by the peak temperature of the system. Panel (a) of Figure
\ref{fig:rmax} shows this fractional temperature drop versus mean
temperature. We can test for possible correlation here using Kendall's rank
order correlation coefficient $\tau$, which is found to be 0.15 with a
p-value (probability of chance occurrence) of 0.38, indicating 
no significant correlation within our sample. However, this is
partly due to the low value of the central temperature decline in
our hottest system, NGC~4073, and excluding this system yields $\tau$ = 0.27
and a p-value of 0.14, indicating a positive correlation at the $>$1$\sigma$
level. The values of the fractional dip in Figure \ref{fig:rmax}(a) are clearly
for the most part lower than the value 0.6 which is typical in
CC clusters \citep[see Fig.7 of][]{sanderson06}, in agreement
with the findings of \citet{rasmussen07}. The CC group with the lowest 
fractional temperature drop (\ensuremath{T_{drop}\,/\,T_{peak} \sim 0.1}) 
is SS2b153, the central galaxy of which is NGC 3411 \citep{mahdavi05}. 
This group has been interpreted as evidence of a CC system that 
has been re-heated by recent AGN activity \citep{osullivan07b}. 

The \Chandra\ study of \citet{rasmussen07} found that the radius at which
the temperature peaks is correlated with group temperature (and hence group
size). We test for a correlation between the physical 
sizes of the CCs and the value of \RF\ (see Figure \ref{fig:rmax}, 
panel (b)) in our sample. The value of Kendall's rank order
correlation coefficient $\tau$ is 0.29, with a p-value of 0.10, so we
have some evidence for the existence of a correlation:
larger systems tend
to have larger CCs, which is perhaps not a surprising result, and
agrees qualitatively with the analysis of \citet{rasmussen07}. Fitting a
linear model using an orthogonal regression \citep{isobe90} to allow for intrinsic scatter,
we find a relationship between the size of the CC and the value of
\RF\ of the following form,
\begin{equation}
r(T_{max})=(0.14\pm0.07)\RF-(1.0\pm15.4)\kpc.
\end{equation}
where \ensuremath{r(T_{max})} is the radius of maximum temperature in kpc,
and \RF\ is also measured in kpc. This relation yields a flatter slope than
the analysis of \citet{rasmussen07}, who found a slope of
0.20$\pm$0.02~\RF, although the large error bars mean that the slopes of
the two relations agree within errors. There are some
differences in the methodology used to estimate
\RF\ in the two studies, with the \citet{rasmussen07} values for the
groups in common tending to be larger, which could contribute to the
slightly different results obtained.

Our result has essentially zero intercept, so that 
$r(T_{max})\approx$~0.14\,\RF.
This is very similar to the ratio of the temperature peak radius to \RF\ 
found for 
clusters by \citet{sanderson06}. Hence it seems that although the 
{\it depth} of the CC is less in groups than clusters, their
{\it size} (scaled to \RF) is similar. In Figure \ref{fig:rmax} panel
(c), we scale the temperature peak radius to \RF\ and find no significant
temperature trend in the {\it scaled} size of the CCs.

Comparing the global properties of the CC and NCC groups
in our sample,
the mean temperature of the NCC systems is found to be
0.89$\pm$0.08\keV, compared to 1.11$\pm$0.07\keV\ for the CC groups. 
Therefore, although the two populations overlap, the centroid of 
the distribution is significantly lower for the NCC subsample.
Similarly, the mean X-ray luminosities of the subsamples are
log \LX\ = 42.70$\pm$0.12~\ergps\ for the CC groups, and 
log \LX\ = 42.20$\pm$0.21~\ergps\ for the NCC systems. 
Of course, we are not dealing with a statistical sample of
groups, so it is important not to over-interpret these results.
However, one might expect selection effects to work in the
opposite direction --- cool, faint groups will be easier to detect
if they have CCs --- so our results provide tentative evidence
that cooler groups are more likely to lack CCs. Moreover, it
is interesting to note that three of the four coolest systems, with
temperatures less than 0.7~keV, do not have CCs, which hints
at the possibility of a lower temperature threshold for CC systems.
Further study of low temperature groups (which are faint, and therefore 
hard to detect) is required to confirm whether there is indeed such a 
threshold.

To summarise, for our sample the fractional central drop in
temperature is smaller than that seen in clusters.
We see no clear trend in CC strength with mean temperature
within our group sample, but the coolest systems ($\bar{T}$\,$<$\,0.7\,\keV)
tend not to show CCs at all. We find a relation rather similar
to that of \citet{rasmussen07} between the
physical size of the CC region (defined as the radius of maximum
temperature) and \RF. Smaller groups (lower \RF) exhibit smaller 
CCs, and the core size across our temperature range is typically
14\% of \RF, in good agreement with what is seen in clusters.
\section{Radial Profiles of Gas Properties}
\label{sec:radial}
Here we look at the radial variation of deprojected gas properties:
entropy, pressure and gas density. These
quantities have been derived from a two dimensional analysis, rather than
an annular approach, so we define the characteristic radius of each
measurement as the mean of the two bounding radii of the spectral extraction
region.
\subsection{Entropy profiles}
\label{sec:entropy}
The entropy of a relaxed system, coupled with knowledge of the potential
well, completely defines the properties of the intracluster medium
\citep[e.g.][]{voit03b}. Gas entropy can provide useful insights
into the non-gravitational processes shaping the ICM, since it
remains unchanged when gas is simply moved around within a system.
We define entropy as \citep[e.g.][]{ponman03},
\begin{equation}
S = T~n_{e}^{-2/3},
\end{equation}
where ${T}$ is the mean temperature of the system, and \ensuremath{n_{e}}
is the gas density, and $S$ has units of keV~cm$^{2}$. This is related to
the true thermodynamic definition of entropy via a logarithm and additive
constant, and has the benefit of acting as a proxy for entropy which follows
directly from two X-ray derived properties.
\begin{figure}
\includegraphics[width=8cm]{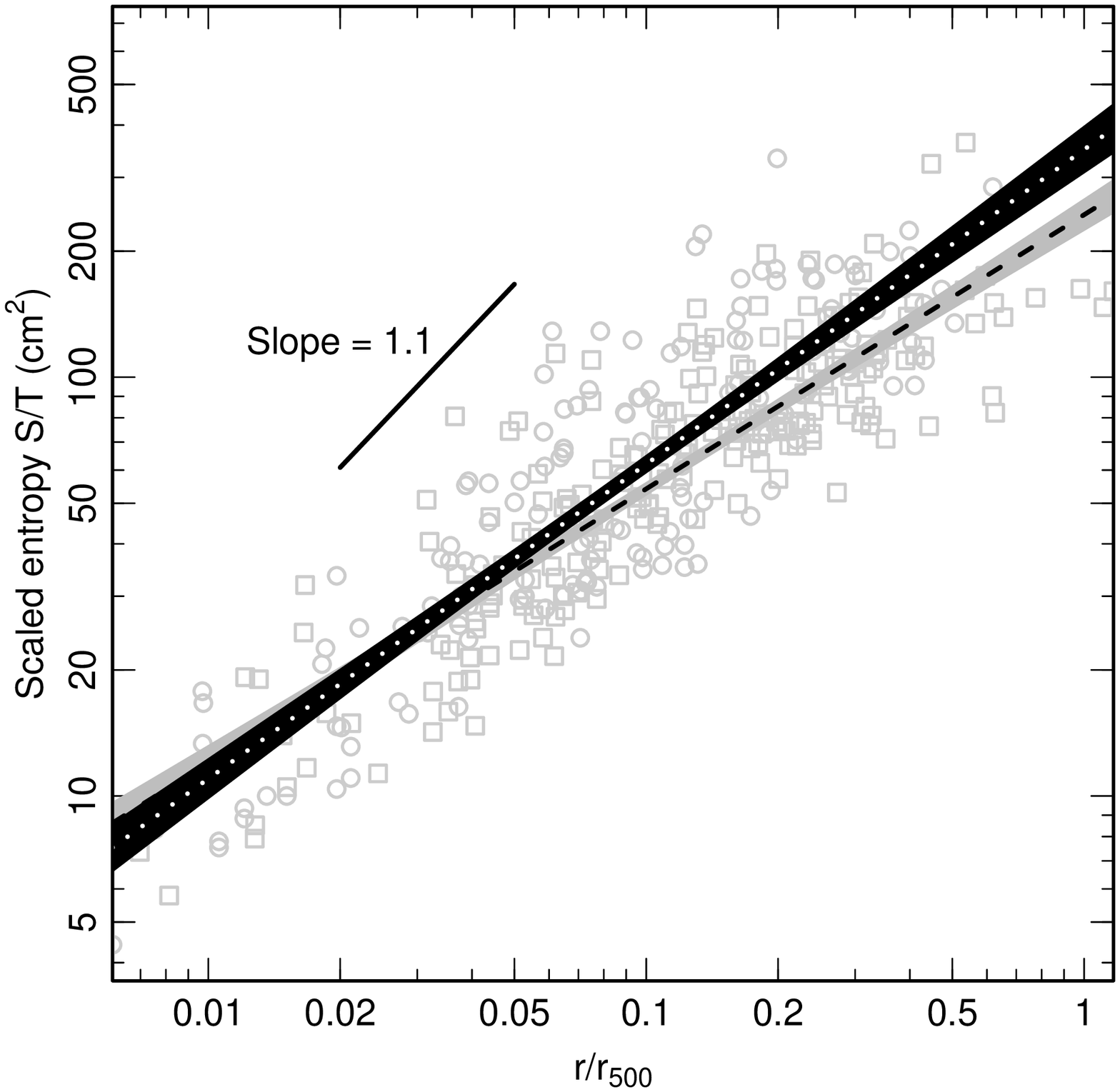}
\caption{Power-law fits to the entropy profiles for all groups in the sample, 
scaled self-similarly, as \ensuremath{S\,/\, \bar{T}}. The white dotted 
line and associated 68\% confidence region show a power-law fit to the 
\textit{Cool} groups, and the black dashed line and associated 68\% 
confidence region show a power-law fit to the \textit{Warm} groups. 
The grey data points show individual entropy measurements for the 
Cool groups (open circles) and the Warm groups (open squares). The
theoretical slope of \ensuremath{r^{1.1}} \citep{tozzi01} showing the
expectation from gravitational processes only is also shown.}
\label{fig:S-T}
\end{figure}

In the self-similar case, entropy simply scales with the virial temperature
of the system. However, a modified entropy scaling of
\ensuremath{S\,/\,\bar{T}^{2/3}} was found to perform well across a wide
range of virial masses, from groups to clusters \citep{ponman03}. We aim
here to test the use of this modified entropy scaling
across the group regime. Considering first the self-similar scaling
(${S}$~$\propto$~$\bar{T}$),
Figure \ref{fig:S-T} shows power-law fits (in log-log space) to these
scaled entropy profiles for both the Cool and Warm sub-samples, with their
associated 68\% confidence regions\footnote{Throughout this section, the
estimation of confidence regions on power-law fits uses a modification of
the `confidence.band' function written for the \rr\ statistical environment
package by Derek Young and David Hunter, see
http://www.stat.psu.edu/$\sim$dhunter/R/confidence.band.r}. We note 
that few groups contribute beyond $\sim$0.5\,\RF, and hence results shown here 
beyond this radius should be treated with caution. If the self-similar
scaling works then it should remove any systematic temperature dependence
in the profiles, so that the fits to the Cool and Warm samples should
coincide. In practice, one can see that outside the core the
scaled profile for Cool groups lies significantly above that for
Warm ones, indicating that a scaling less strong than 
${S}$~$\propto$~$\bar{T}$ is required.

\subsubsection{Optimum entropy scaling}
\label{S:scale}
\begin{figure}
\includegraphics[width=8cm]{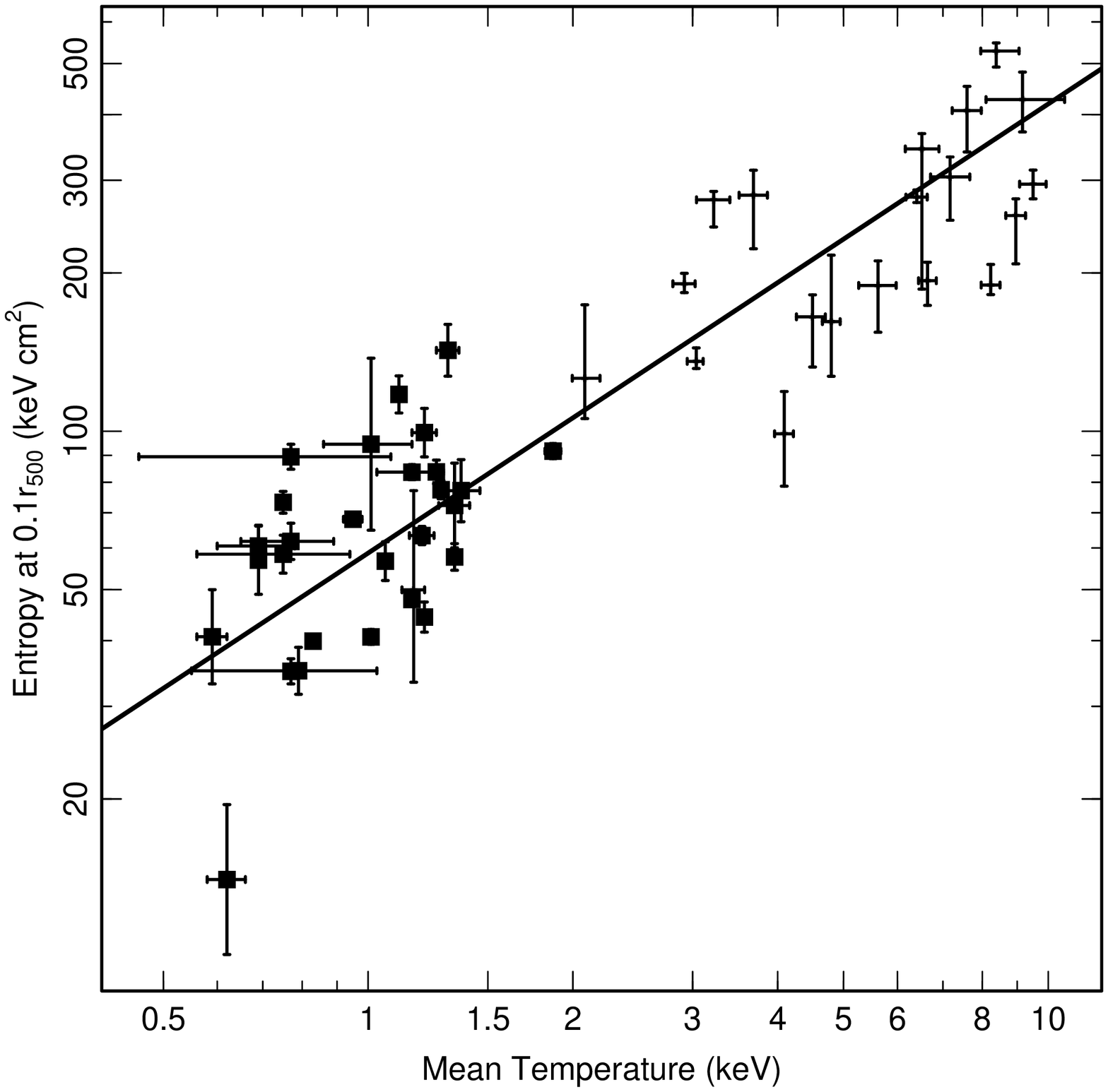}
\caption{Entropy at 0.1\,\RF\ as a function of the mean temperature of the system (measured between 0.1 and 0.3\,\RF) for the 2dXGS sample (solid squares). The cluster sample of \citet{sanderson08} is shown as simple error bars. The solid line shows a BCES orthogonal regression fit \citep{akritas96} performed in log-log space to the combined group and cluster sample.}
\label{fig:Sscale}
\end{figure}
We can investigate the optimum entropy scaling by looking at the entropy at
a particular radius as a function of temperature. To each individual
(unscaled) entropy profile, we fitted a power law model, to
determine the entropy at 0.1\,\RF. In Figure \ref{fig:Sscale} 
we plot the entropy at this characteristic radius as a 
function of the mean temperature of the system.

We fit a straight line in log-log space to yield the optimum entropy scaling, 
which will simply be 
the slope of this relation. Due to the measurement errors on the mean 
temperature, and the errors on the entropy measurements, which are 
propagated from the errors on the power-law fits to each individual group, 
we fit the relation using the BCES generalisation of the 
orthogonal regression method presented by \citet{akritas96}. This allows 
for both measurement errors (which may be correlated) and also for unknown 
intrinsic scatter in the relation. The narrow temperature range covered 
by the groups, coupled with the large dispersion in group properties, results
in a large error in the slope of the $S$--$T$ relation. We therefore combine
our sample with the cluster data of \citet{sanderson08}  and fit the 
relation using the combined group and cluster sample, to increase the 
temperature baseline.

As described by \citet{akritas96} and \citet{isobe90}, the analytic form 
presented for the estimates of variance on the slope and intercept are only 
suitable for large samples, and more appropriately a bootstrap error analysis 
should be applied to describe the error behaviour of small samples. We have 
calculated both the analytical variance from \citet{akritas96}, and the result 
from a bootstrap analysis. We extracted 100 bootstrap samples from our 
original data, and calculated the slope and intercept in each case, 
determining the mean and standard deviation from these results. Applying 
the orthogonal regression method yields a slope of 0.79$\pm$0.06 using the 
analytical variance; the slope 
resulting from our bootstrap analysis was found to be 0.78$\pm$0.06. 

The recovered slopes are a little steeper than the 2/3 scaling of
\citet{ponman03}, although they are consistent with it
at the 2\,$\sigma$ level. Our result is in good agreement
with recent work of \citet{sun08}, who find a slope
of 0.78$\pm$0.12 at the somewhat larger radius of 0.15\,\RF, when applying a 
BCES y on x regression to their group sample.
Since the $S\,\propto\,T^{2/3}$ modified entropy scaling of \citet{ponman03}
provides an acceptable representation of the entropy scaling from both
the present study, and the \Chandra\ study of \citet{sun08},
we adopt this scaling for the remainder of this paper. However, we
should bear in mind that the actual scaling now seems likely to be rather
steeper than $\bar{T}^{2/3}$.

\subsubsection{Cool and Warm samples}
\begin{figure}
\includegraphics[width=8cm]{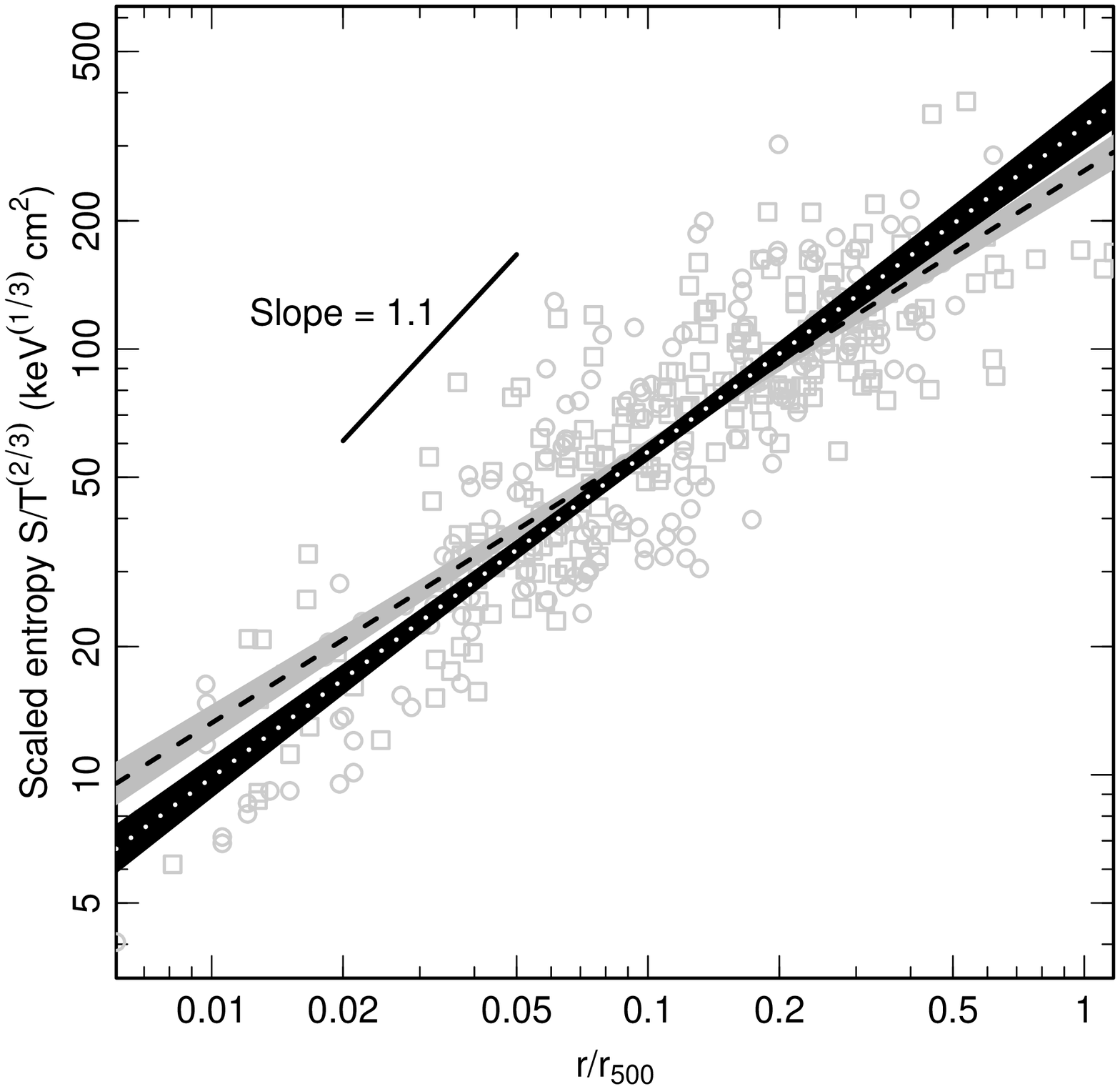}
\caption{Power-law fits to the scaled (\ensuremath{S\,/\,\bar{T}^{2/3}}) entropy profiles for all groups in the sample. The white dotted line and associated 68\% confidence region shows a power-law fit to the \textit{Cool} groups, and the black dashed line and associated 68\% confidence region shows a power-law fit to the \textit{Warm} groups. The grey data points show individual scaled entropy measurements for the Cool groups (open circles) and the Warm groups (open squares). The theoretical slope of \ensuremath{r^{1.1}} \citep{tozzi01} expected from gravitational processes alone is also shown. }
\label{fig:Sprofiles}
\end{figure}
Adopting the modified entropy scaling as discussed above, to remove
the mean effects of group temperature on the entropy, we work with
the scaled entropy \ensuremath{S\,/\,\bar{T}^{2/3}}, which has units
of keV$^{1/3}$~cm$^{2}$. Power-law models are fitted to radial 
profiles of the scaled entropy for the Cool and Warm subsamples
separately. The resulting power-law fits and their error envelopes are
shown in Figure \ref{fig:Sprofiles}. 
Where the entropy is only driven by gravitational processes, its radial
variation is expected to be proportional to \ensuremath{r^{1.1}}
\citep{tozzi01}. Figure \ref{fig:Sprofiles} shows the observed profiles
to be much flatter than this, as also found by
\citet{sun08}. This is consistent with observations of
the inner regions of NCC clusters, although CC
clusters are found to lie closer to the \ensuremath{r^{1.1}} relation
\citep{sanderson08}. 
The slopes of the entropy profiles fitted to the
Cool and Warm groups are 0.77$\pm$0.03 and 0.65$\pm$0.02 respectively,
indicating a steeper slope in the case of the cooler systems. Figure
\ref{fig:Sprofiles} also shows the 68\% confidence bounds on the fits,
arising from errors in the slope and intercept. It can be seen that outside
the core, the scaled entropy for the Cool systems lies somewhat above that
for the Warm groups, confirming the discussion from the previous section, that
the entropy scaling is actually rather steeper than $S\,\propto\,\bar{T}^{2/3}$.

\subsubsection{Core properties}
\label{S:props}
\begin{figure}
\includegraphics[width=8cm]{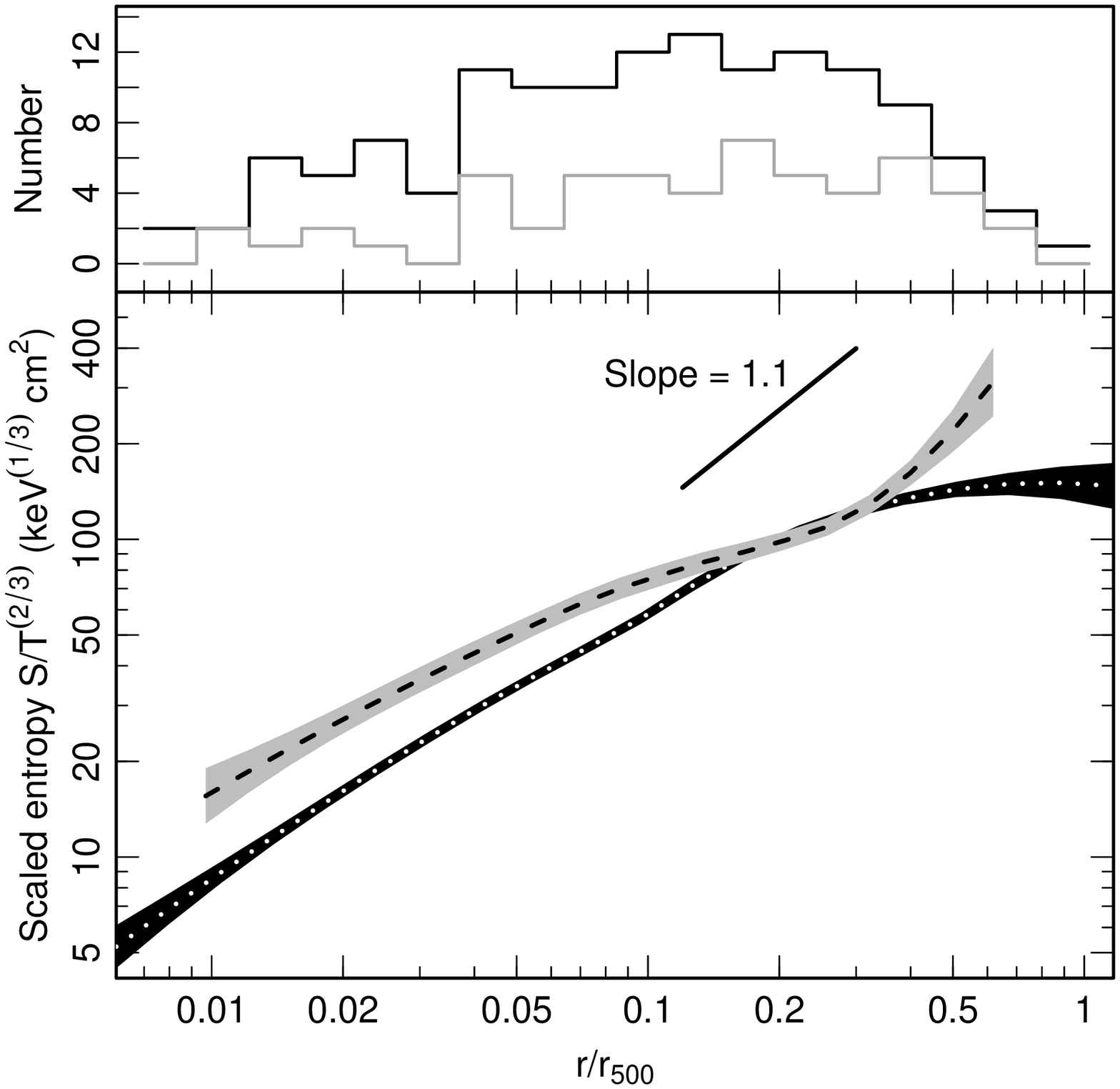}
\caption{Locally weighted regression fits to the scaled entropy profiles of the CC (white dotted line) and the NCC (black dashed line) groups. The confidence regions show the standard errors on the fit. There is no evidence for the need of an entropy pedestal \citep[e.g.][]{donahue06} to describe the CC profile at small radii. The top panel shows the number of groups contributing to the entropy profile fit as a function of radius for the CC (black) and NCC (grey) samples. The behaviour outside 0.5\RF\ is driven by a minority of systems.}
\label{fig:S-full}
\end{figure}
We now consider the entropy profiles of the groups, divided into the
CC and NCC samples described in Section
\ref{sec:divide}. For a sample of CC clusters,
\citet{donahue06} find that the addition of a constant entropy
pedestal of $\sim$10~keV~cm$^{2}$ to a power-law entropy profile
provides a better fit to their data than a simple power-law model. We
use the notation $S^{\prime}$ to refer to scaled entropy,
i.e. $S^{\prime}$ is equivalent to
\ensuremath{S\,/\,\bar{T}^{2/3}}. The pedestal modification is of the
following form,
\begin{equation}
\label{eqn:ped}
S^{\prime}(r)=S^{\prime}_{0}+S^{\prime}_{500}\left(\frac{r}{\RF}\right)^{\alpha},
\end{equation}
where \ensuremath{S^{\prime}_{0}} is the entropy pedestal,
\ensuremath{S^{\prime}_{500}} is the normalisation of the profile at
\RF, and $\alpha$ is the index of the fit. The entropy pedestal
modification of \citet{donahue06} is apparent within $\sim$10~kpc; the
corresponding radial range in our data is
$\sim$\,0.02--0.04\,\RF. However, within this radius there does not
appear to be a flattening of the entropy profiles. This is exhibited
in Figure \ref{fig:S-full}, which shows (unweighted) local regression
fits to the entropy profiles of the CC and NCC
groups. We employed the `\textsc{loess}' algorithm in \rr\ to perform
the fit, and fitted the data in log-log space. The profiles of the
CC groups continually decrease inwards towards a scaled entropy
of $\sim$ 8~keV$^{1/3}$~cm$^{2}$ at radii within 0.01\,\RF. However,
these fits are unweighted, so may be adversely affected by outlying
points as their measurement errors are not accounted for. Instead, we
directly fit the pedestal model of Equation \ref{eqn:ped} to determine
if there is a statistical reason to accept a pedestal addition to the
profile. We find the following best-fitting relation when applying a
weighted (with weights equal to the inverse measurement variance at
each point) non linear regression fit, where the errors quoted are
1\,$\sigma$:
\begin{equation}
S^{\prime}(r)=0.3~(\pm1.3)+285~(\pm17)\left(\frac{r}{\RF}\right)^{0.79~(\pm0.04)},
\end{equation}
suggesting the pedestal modification is not statistically required, as
$S^{\prime}_{0}$ is consistent with zero at the level of
1\,$\sigma$. Numerical simulations that employ only radiative cooling
and the effects of gravity have shown the value of the entropy
pedestal in CC clusters to tend to zero as the system ages
\citep{ettori08}. In the coolest system considered (4~keV) by
\citet{ettori08}, this occurs in approximately 2~Gyr, and the entropy
profile of the system is also seen to steepen during this
time. Unfortunately, we do not have enough radial coverage for all
groups to test the success of a pedestal model on individual group
profiles.

We can also see a suggestion of flattening in the entropy profiles of
CC groups at larger radii, between 0.5\,\RF\ and \RF\ (see Figure
\ref{fig:S-full}). There is also an apparent up-turn in the NCC profile
at large radius. The top panel of Figure \ref{fig:S-full} shows the
number of groups contributing at each radius. The effect seen in the CC
profile at large radius is driven solely by the group RGH~80 as the radial 
coverage for the majority of CC groups extends only as far as 0.5\,\RF\ 
(five groups each add one data point at radii larger than 0.5\,\RF). 
Similarly, the
apparent up-turn in the NCC profile is driven by only 6 data points
contributing outside 0.35\,\RF, and hence this feature should be
regarded with caution. Therefore, comparing the profiles in Figure 
\ref{fig:S-full}, the greatest difference between the CC and NCC profiles
occurs within 0.15\,\RF, where the NCC systems show higher entropy levels than
the CC systems. At approximately 0.15\,\RF\ the profiles appear to converge,
ignoring the artifacts introduced by the fits at large radius, as described 
above.
\begin{figure}
\includegraphics[width=8cm]{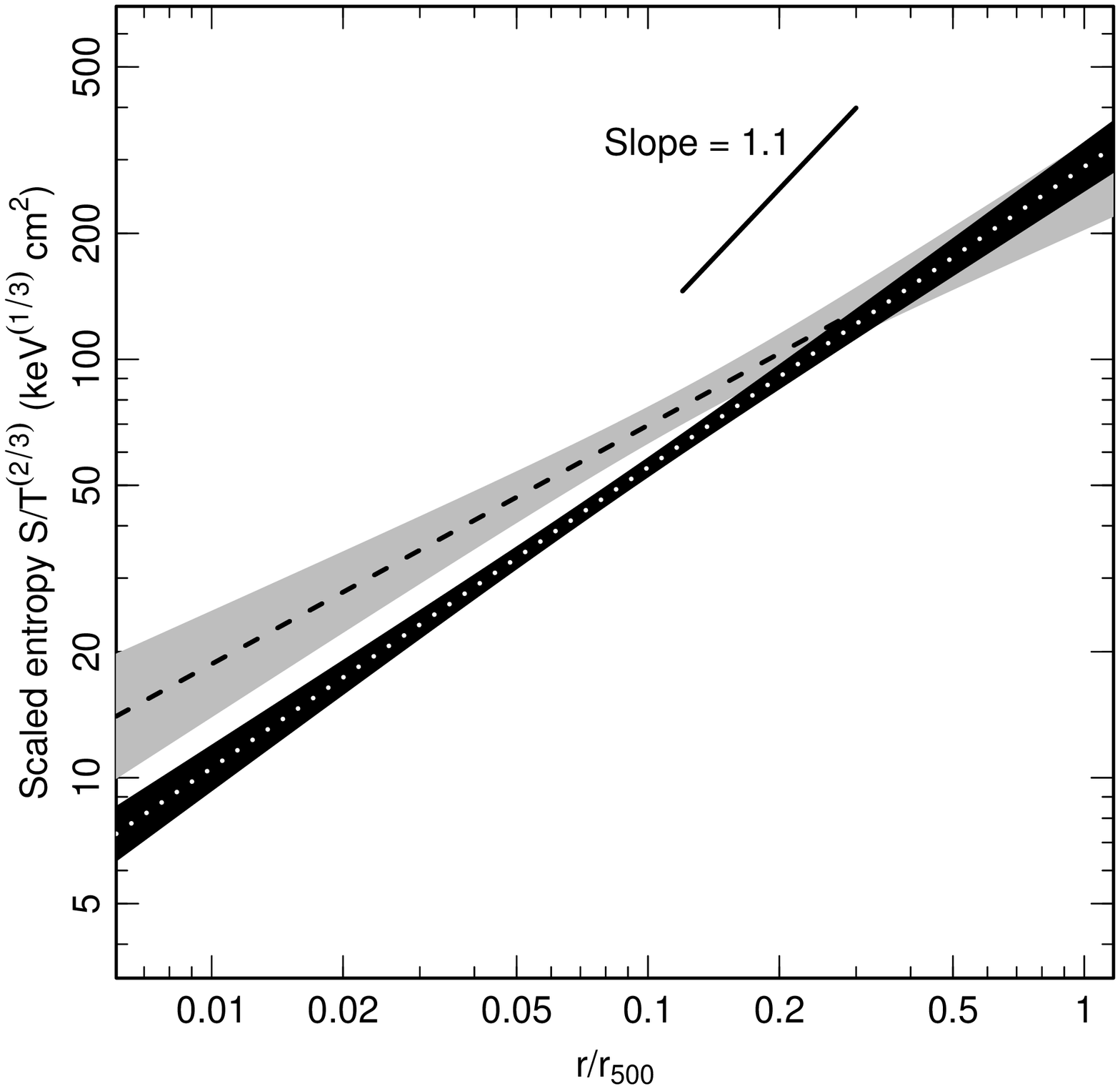}
\caption{Power-law fits to the entropy profiles of the groups in the sample, divided into the CC (white dotted line with black 95\% confidence region) and NCC (black dashed line with grey 95\% confidence region) groups. The theoretical slope of \ensuremath{r^{1.1}} \citep{tozzi01} is also shown.}
\label{fig:SprofCCNCC}
\end{figure}

As we are interested in comparing the slopes of the CC and NCC samples directly, we proceed in Figure \ref{fig:SprofCCNCC} to fit power-law models in log-log space to the CC and NCC samples separately. These take the form,
\begin{equation}
\label{eqn:pl}
S^{\prime}(r)=S^{\prime}_{500}\left(\frac{r}{\RF}\right)^{\alpha}
\end{equation}
which allows the derivation of the slope of the entropy profiles,
allowing a comparison to that expected from gravitational effects
alone. As already indicated in the locally weighted regression fits
used in Figure \ref{fig:S-full}, we see a significant difference in
the observed slopes of the entropy profiles of the two samples. Figure
\ref{fig:SprofCCNCC} shows the 95\% confidence bounds on the power-law
fits, and at this level, the power-law fit to the NCC groups is
flatter than that of the CC groups, within a radius of
$\sim$0.15\,\RF. The fitted slope to the CC systems is 0.71$\pm$0.02,
compared to 0.57$\pm$0.04 for the NCC groups, both flatter than the
expectation for gravitational effects alone \citep{tozzi01}. Outside
$\sim$0.15\,\RF, the two power-law fits begin to converge,
suggesting that the processes affecting the core have little effect
at larger radii. As we will further show in Section \ref{sec:S-ind},
the NCC groups also show considerably more scatter in their entropies
at small radii, evidenced by the broad confidence region in Figure
\ref{fig:S-full}, compared to their CC counterparts.
 
\subsubsection{Individual groups}
\label{sec:S-ind}
\begin{table}
  \centering
  \caption{The results of the power-law fits to the scaled entropy profiles of individual groups. Shown are the fitted index, and the scaled entropy at 0.01\,\RF, 0.1\,\RF\ and 0.5\,\RF. The result for fitting a power-law to all the groups simultaneously is also shown.}
  \label{table:Sprops}
  \begin{tabular}{ccccc}
  \hline
     Group & Index & S$_{0.01\RF}$ & S$_{0.1\RF}$ & S$_{0.5\RF}$\\
  \hline
  All & 0.70 & 11.4 & 57.7 & 179\\
  \hline
  3C 449 & 0.41 & 24.9 & 64.6 & 126\\
  A 194 & 0.36 &  47.6 & 108.1 & 192\\
  HCG 15 & 1.32 & 1.6 & 33.7 & 282\\
  HCG 42 & 1.02 & 9.5 & 99.2 & 510\\
  HCG 51 & 0.66 & 17.2 & 78.1 & 225\\
  HCG 62 & 0.83 & 8.6 & 57.9 & 220\\
  HCG 68 & 0.77 & 15.6 & 92.8 & 323\\
  HCG 92 & 1.12 & 4.6 & 60.5 & 369\\
  HCG 97 & 0.54 & 16.9 & 58.2 & 138\\
  IC 1459 &  0.45. & 25.0 & 71.2 & 148\\
  NGC 507 & 0.70 & 9.9 & 49.4 & 152\\
  NGC 533 & 0.72 & 13.1 & 68.3 & 217\\
  NGC 2300 & 1.14 & 8.3 & 113.0 & 703\\
  NGC 2563 & 0.72 & 21.5 & 113.9 & 366\\
  NGC 4073 & 0.67 & 13.6 & 64.0 & 189\\
  NGC 4168 & 0.79 & 17.1 & 105.4 & 377\\
  NGC 4261 & 0.57 & 31.5 & 118.1 & 297\\
  NGC 4325 & 0.93 & 4.7 & 39.6 & 177\\
  NGC 4636 & 0.69 & 8.1 & 39.9 & 122\\
  NGC 5044 & 0.81 & 7.2 & 46.5 & 171\\
  NGC 5129 & 0.75 & 12.6 & 71.0 & 237\\
  NGC 5171 & 0.22 & 60.6 & 100.1 & 142\\
  NGC 5846 & 0.92 & 10.6 &  86.7 & 378\\
  NRGb 184 & 0.73 & 12.7 & 67.7 & 218\\
  Pavo &  0.55 & 18.5 & 65.9 & 160\\ 
  RGH 80 & 0.56 & 12.8 & 47.1 & 117\\
  SRGb 119 & 1.09 & 6.3 & 78.3 & 455\\
  SS2b153 & 0.73 & 8.6 & 46.7 & 152\\
  \hline
  \end{tabular}
\end{table}

To look at the properties of individual groups, and how they relate to
the mean trends, we fitted the scaled entropy profile for each group
with a power-law model. In addition to the index of the fit, we
determined the value of the entropy at three characteristic radii:
0.01\,\RF, 0.1\,\RF, and 0.5\,\RF. The values of each of these
parameters is shown in Table \ref{table:Sprops}. We also fitted a
power-law to the combined data from all groups, which yielded a slope
of 0.70$\pm$0.02. The slope and entropy measurements for each group
indicate how each group varies from the mean, or expected theoretical,
trends. NCC clusters are seen to show more scatter in their
entropy profiles at small radii \citep{mccarthy08}. This trend is also
seen in our group sample, as for CC groups the rms scatter in scaled
entropy at 0.01\,\RF\ is 6~keV$^{1/3}$~cm$^{2}$, whereas for
NCC groups it is three times larger: 18~keV$^{1/3}$~cm$^{2}$. 
In contrast, at 0.1\,\RF, the profiles
are much more comparable. The mean scaled entropy at 0.1\,\RF\ for the
CC groups is 70~keV$^{1/3}$~cm$^{2}$, with an rms scatter of 
26~keV$^{1/3}$~cm$^{2}$, while for NCC systems the corresponding figures 
are 78~keV$^{1/3}$~cm$^{2}$ and 23~keV$^{1/3}$~cm$^{2}$. This lends
more support to the observation in Section \ref{S:props} that the
CC and NCC entropy profiles converge outside 0.15\,\RF.

The fitted parameters for Hickson 15 show it to have a relatively
steep entropy profile, with very low central entropy. These are
unexpected properties for NCC groups. Conversely, the fit to
NGC\,5171 shows a very flat slope and high central entropy. However,
both these groups are current mergers, and hence likely to be far
from equilibrium. NGC\,5171 has been studied in detail by \citet{osmond04}, 
and HCG\,15 shows signs of having undergone a recent merger (E. O'Sullivan, private communication). Moreover in both these groups, the
two-dimensional data cover only a small
radial range, and the entropy fit has to be extrapolated well beyond the
data to provide the values at the inner and outer radii tabulated in
Table~\ref{table:Sprops}. Anomalous results for these two systems are
therefore not unexpected.

\subsection{Pressure profiles}
\label{sec:P-overall}
Applying the modified entropy scaling of \citet{ponman03}, the gas density
scales as $\rho \propto \bar{T}^{1/2}$, resulting in a
scaling for pressure of \ensuremath{P~\propto~\bar{T}~^{3/2}}. To
compare across the group sample, we therefore scaled the pressure profiles
by $\bar{T}^{\,-3/2}$.

In an analogous manner to the analysis of the entropy profiles, we
again divided the groups into the Cool and Warm samples, to see if
there were any systematic differences in the pressure profiles. F+06
found that fitting a power-law to the pressure profiles did not yield
a good fit, and they instead followed the approach of
\citet{sanderson05} in fitting a locally weighted regression curve to
the data. This has the advantage of yielding a fit to the data that is
independent of any model assumptions, and which also suppresses
scatter. We follow the same approach here, and fit a non-parametric
curve to the data (in log-log space) using the \rr\ function `\textsc{loess}'.
\begin{figure}
\includegraphics[width=8cm]{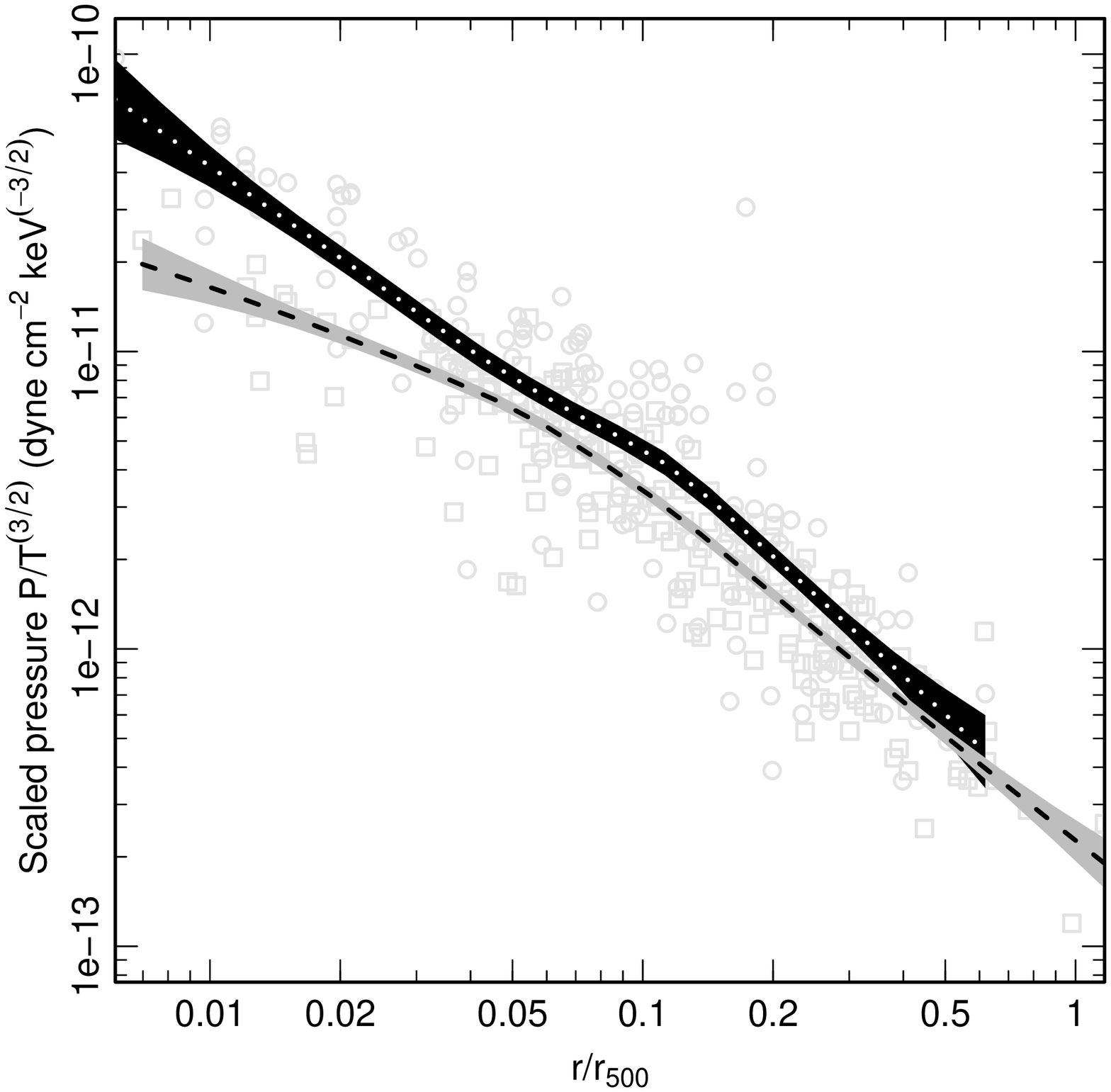}
\caption{A comparison of the pressure profiles between Cool (dotted white line) and Warm (dashed black line) groups. The curves are locally weighted regression fits to the Cool groups and Warm groups separately. The confidence regions show the standard errors associated with the fit. The grey data points show individual scaled pressure measurements for the Cool groups (open circles) and the Warm groups (open squares).}
\label{fig:Pprofs}
\end{figure}

Figure \ref{fig:Pprofs} shows the fitted pressure profiles to the Cool
and Warm group samples. The cooler groups are seen to exhibit steeper
pressure profiles in the inner regions, most prominently within
approximately 0.05\,\RF, rising to a larger value of scaled pressure
at the innermost radius. Outside $\sim$\,0.15--0.2\,\RF, the fits to the
two sub-samples begin to converge. It is possible that we are seeing
the effect of the brightest group galaxy on the pressure
profiles here, as the difference is most apparent within a small
radius. The characteristic size of the BGG can be estimated using \D,
the diameter of the isophote where surface brightness in the $B$-band
is 25~mag/arcsec$^{2}$. We converted values of \D\ from the RC3 galaxy
catalogue \citep{RC3} into physical radii using the distance
information given in Tables \ref{table:main} and \ref{table:main2}. We
find the mean size of the BGGs to be 0.05$\pm$0.02\,\RF. Separating
this into the Cool and Warm samples yields the same result, i.e. there
is no systematic difference in the sizes of the central galaxies as a
fraction of \RF. We postulate that the difference in pressure
therefore arises from a difference in the dominance of the stellar
mass component in the BGGs of Cool and Warm groups.

We can examine this idea by comparing the masses of the group cores, and 
comparing with the luminosities of their central 
galaxies. Assuming hydrostatic equilibrium, the total group mass within radius $r$
is given by,
\begin{equation}
\label{eqn:mass}
M(r)=-\frac{r^{2}}{G\rho}\frac{dP}{dr},
\end{equation}
where $P$ is the pressure and $\rho$ is the density of the gas. We
evaluated Equation \ref{eqn:mass} for the Cool and Warm samples
separately, using the `canonical' fits to the pressure (Figure
\ref{fig:Pprofs}) and gas density profiles (see Figure
\ref{fig:rhoprofs}), and removing the applied temperature
scaling. Within 0.05\,\RF\ (roughly the optical extent of a typical
BGG), the masses of the Cool and Warm systems are
2.1$\times$10$^{11}$\Msol\ and 3.7$\times$10$^{11}$\Msol\
respectively. Taking into account the difference in \RF\ between the
Cool and Warm groups, these recovered masses are typical
of brightest group galaxies at this radius \citep[e.g.][]{humphrey06}.

We can test the stellar dominance by calculating the mass-to-light
ratio of the Cool and Warm systems. The mean $B$-band BGG
luminosities (with standard errors) for the Cool and Warm samples are
4.9$\pm$0.9$\times$10$^{10}$\,$L_{B,\odot}$ and
5.7$\pm$0.9$\times$10$^{10}$\,$L_{B,\odot}$ respectively. Using the above 
masses then gives $B$-band mass-to-light ratios, within 0.05\,\RF, of 
4.3 and 6.5 for the Cool and Warm groups
respectively. This indicates increased dominance from
the stellar mass in the cooler groups, which would lead to a more
concentrated mass profile, and hence an excess pressure relative to
the warmer groups.
\begin{figure}
\includegraphics[width=8cm]{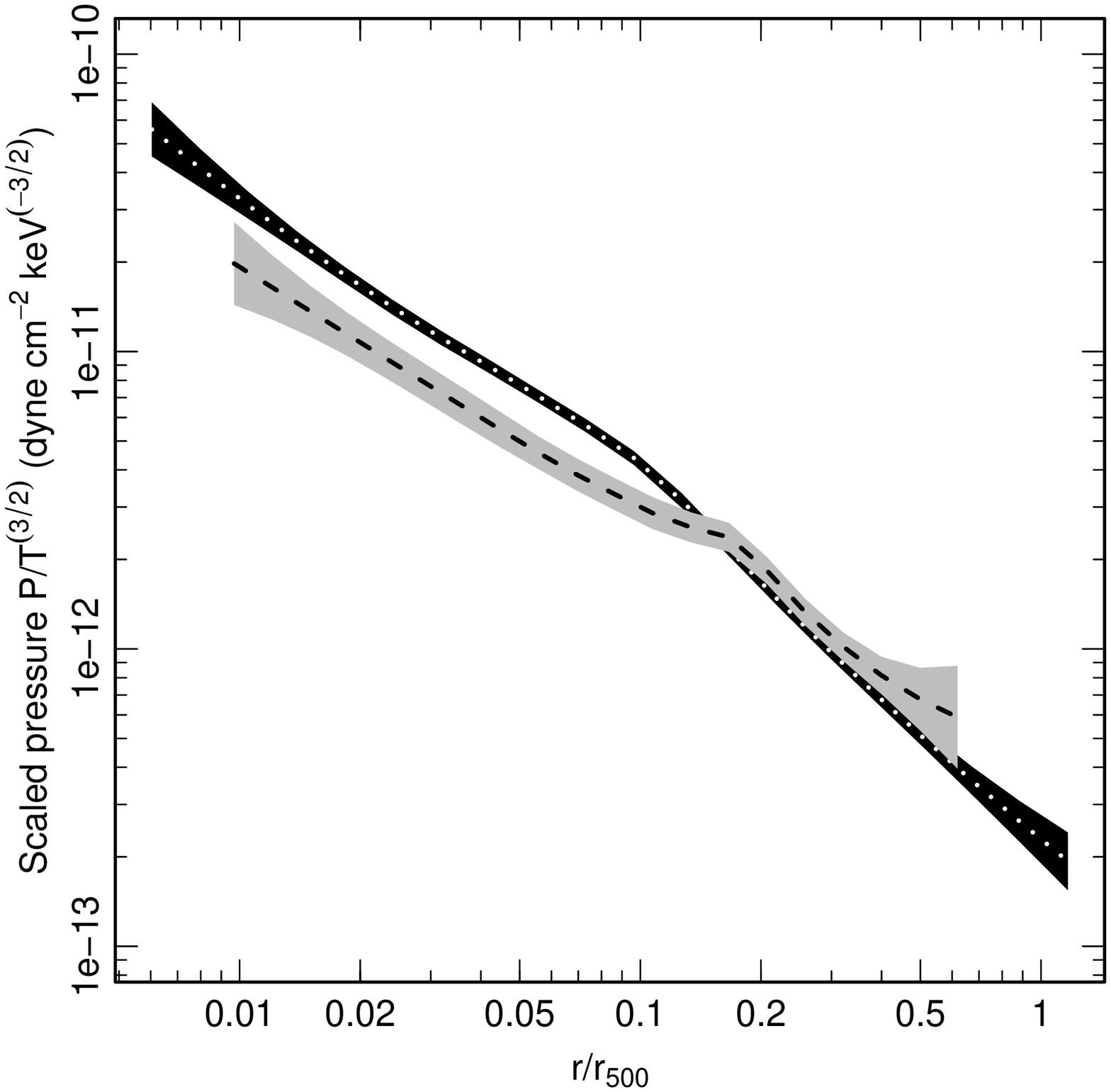}
\caption{A comparison of the pressure profiles between CC (dotted white line) and NCC (dashed black line) groups. The curves are locally weighted regression fits to the CC and NCC groups separately. The confidence regions show the standard errors associated with the fit.}
\label{fig:Pprofs-cc}
\end{figure}

Dividing the sample into the CC and NCC classes, the same fitting
procedure can be applied, and the fitted pressure profiles of the
sample split on the basis of their core properties are shown in Figure
\ref{fig:Pprofs-cc}. The CC groups exhibit somewhat
higher pressure than the NCC groups, although the two profiles overlap
at $\sim$~0.15\,\RF. The typical radius of maximum temperature was
found to be $\sim$\,0.14\,\RF\ (see Section \ref{sec:core_class}), so
the excess pressure is seen only within the region of the CC. 
This is to be expected, since hydrostatic equilibrium (equation 7) 
gives a steeper pressure gradient (for the same $M(r)$) when gas density
is higher, as it is within CCs.

\subsection{Gas density profiles}
\label{sec:density}
We also attempt to draw conclusions on the systematic differences
between groups by examining the gas density profiles of the
sample. The modified entropy and pressure scaling applied to the group
sample so far implies a scaling for gas density of
\ensuremath{\rho~\propto~\bar{T}^{1/2}}, which we apply to the sample.

\begin{figure}
\includegraphics[width=8cm]{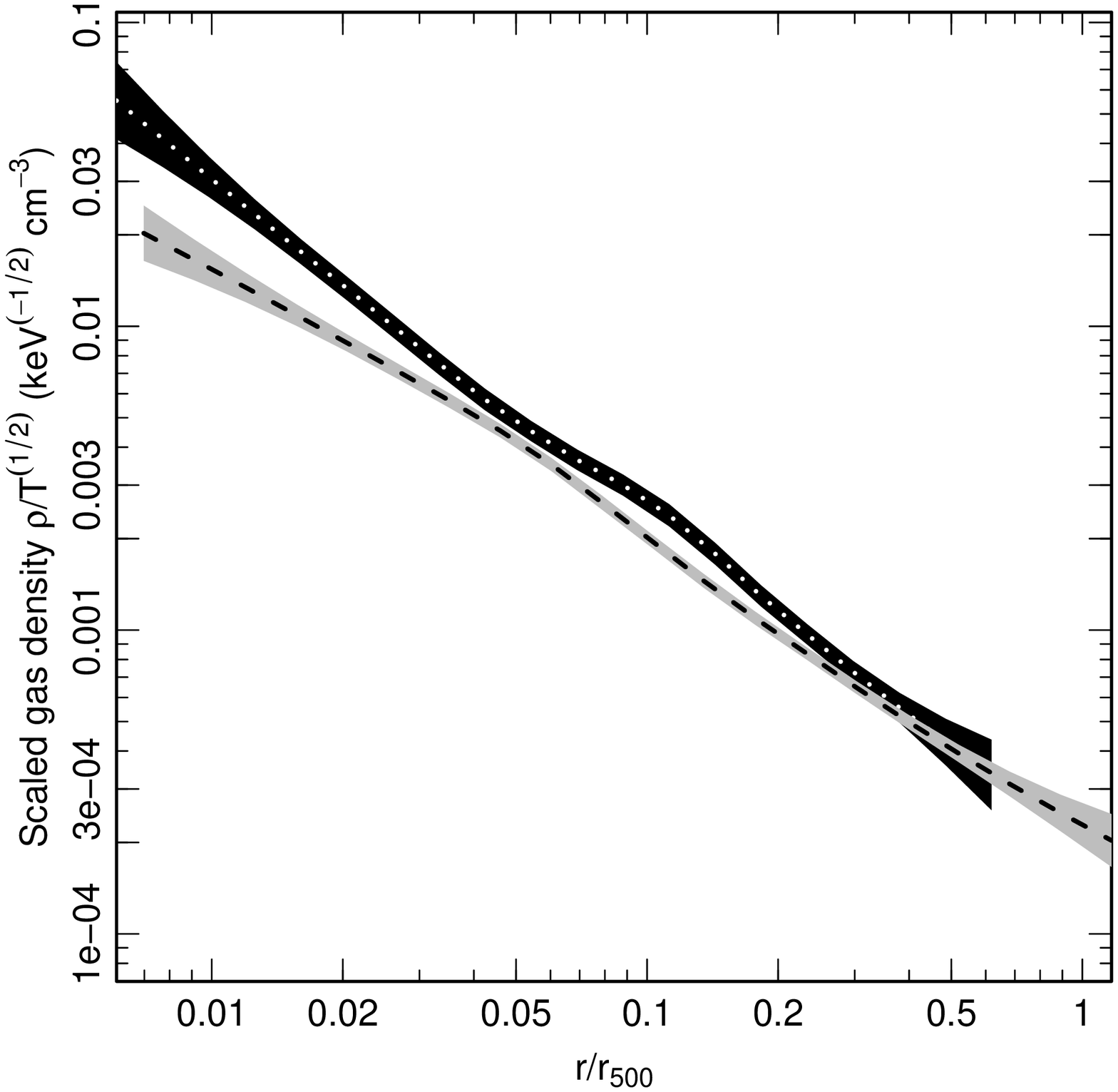}
\caption{A comparison of the gas density profiles of Cool (white dotted line) and Warm (black dashed line) groups. The curves are locally weighted regression fits to the data, and the confidence regions show the standard errors on the fit.}
\label{fig:rhoprofs}
\end{figure}

We follow an analogous procedure to examining the pressure profiles of
the systems in examining the gas density profiles of the Cool and Warm
systems. We performed locally weighted regression fits on the gas
density profiles of the Cool and Warm groups separately, in log-log
space, using the \rr\ \textsc{`loess'} algorithm. The results of these
fits are shown in Figure \ref{fig:rhoprofs}. The gas density profiles
of the Warm and Cool groups again appear to diverge at approximately
0.05\,\RF, with the Cool groups showing a higher scaled central
density compared to the Warm groups. As explained in Section
\ref{sec:P-overall}, this is showing the effect of the central
galaxy. The pressure of the Cool systems is more strongly peaked, and
consequently, the gas density is affected in the same manner.

\begin{figure}
\includegraphics[width=8cm]{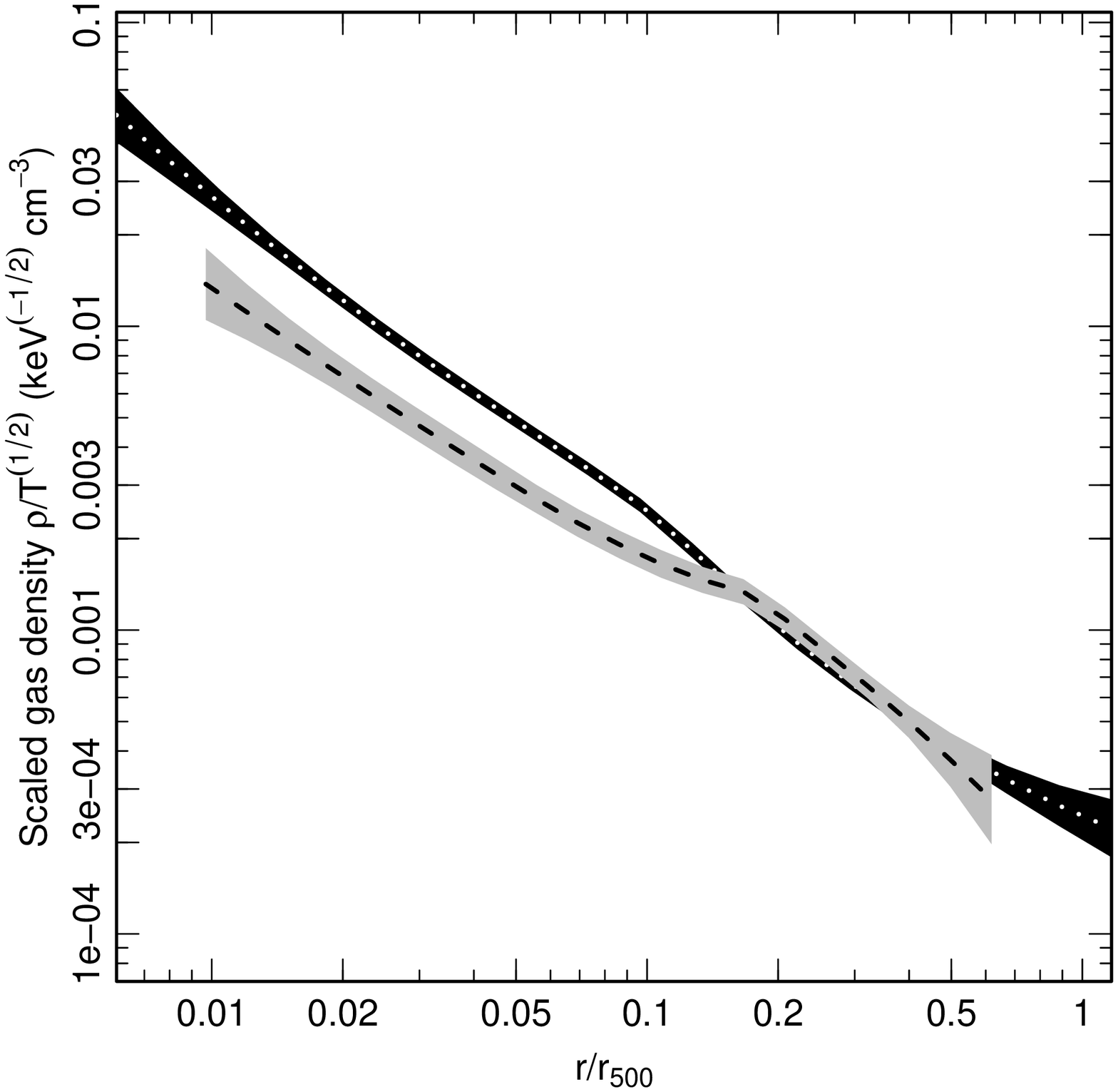}
\caption{A comparison of the gas density profiles of CC (white dotted line) and NCC (black dashed line) groups. The curves are locally weighted regression fits to the data, and the confidence regions show the standard errors on the fit.}
\label{fig:rhoprofs-cc}
\end{figure}

The same comparison can be made dividing the sample on the core
properties of the groups, and this is shown in Figure
\ref{fig:rhoprofs-cc}. In line with the trends seen in the pressure
profiles, the scaled density profiles of CC groups are higher
than those of NCC groups, within a radius of
$\sim$\,0.15\,\RF. This is showing the effects of radiative cooling in
the CC systems, leading to a higher density and pressure in the
region of the core (see Section \ref{sec:P-overall}).
\section{Entropy Distributions}
\label{sec:Sdist}
\begin{figure}
\includegraphics[width=8cm]{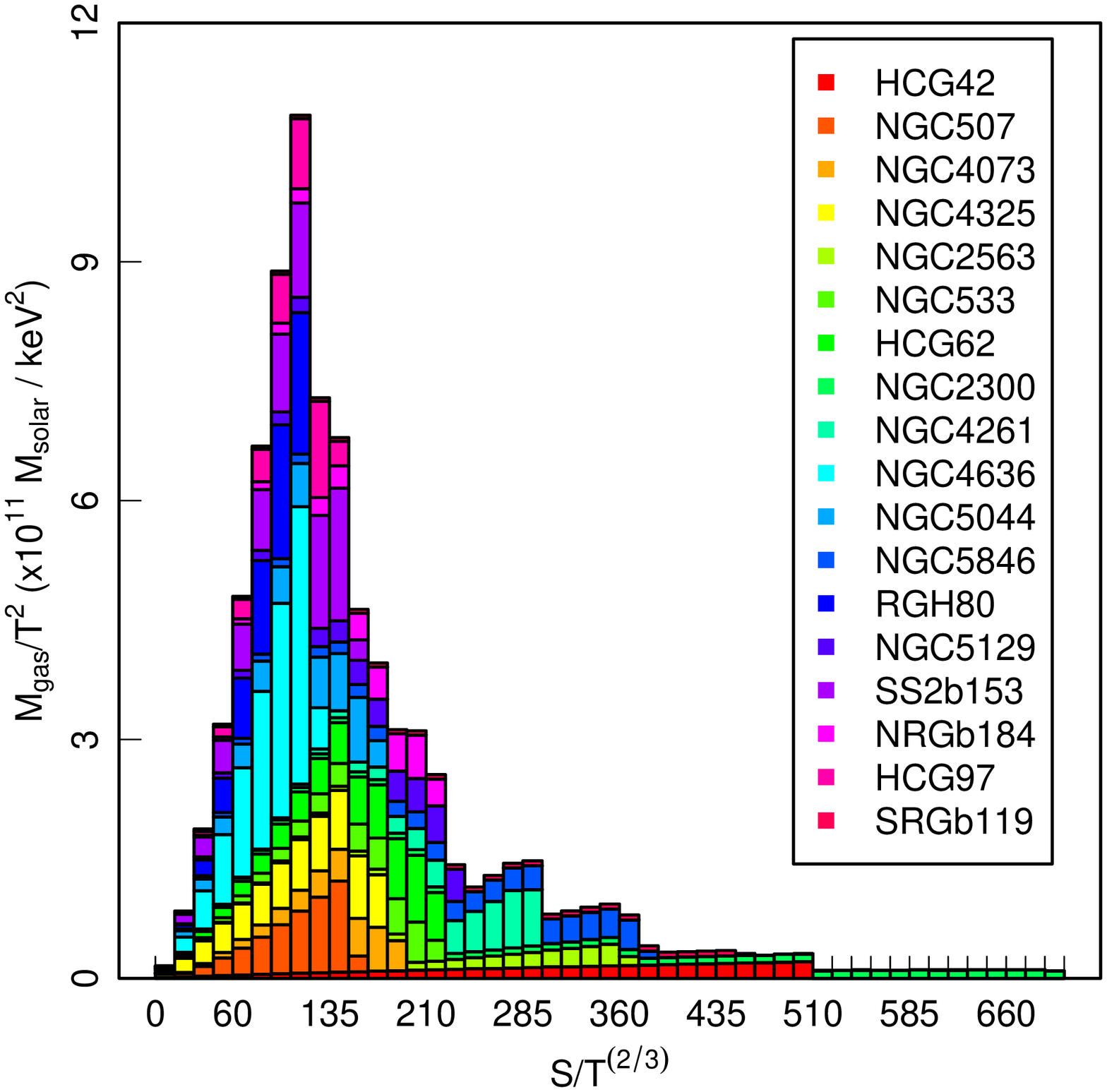}
\caption{Histogram of gas mass as a function of entropy for the CC groups, with each group colour-coded as shown in the legend. This shows the total gas mass in a particular entropy bin, indicating where the majority of the gas mass lies. We have used the modified entropy scaling of $\bar{T}^{2/3}$ to directly compare the systems, and have scaled the gas mass by $\bar{T}^{2}$ (see text for details). }
\label{fig:Shist-CC}
\end{figure}
\begin{figure}
\includegraphics[width=8cm]{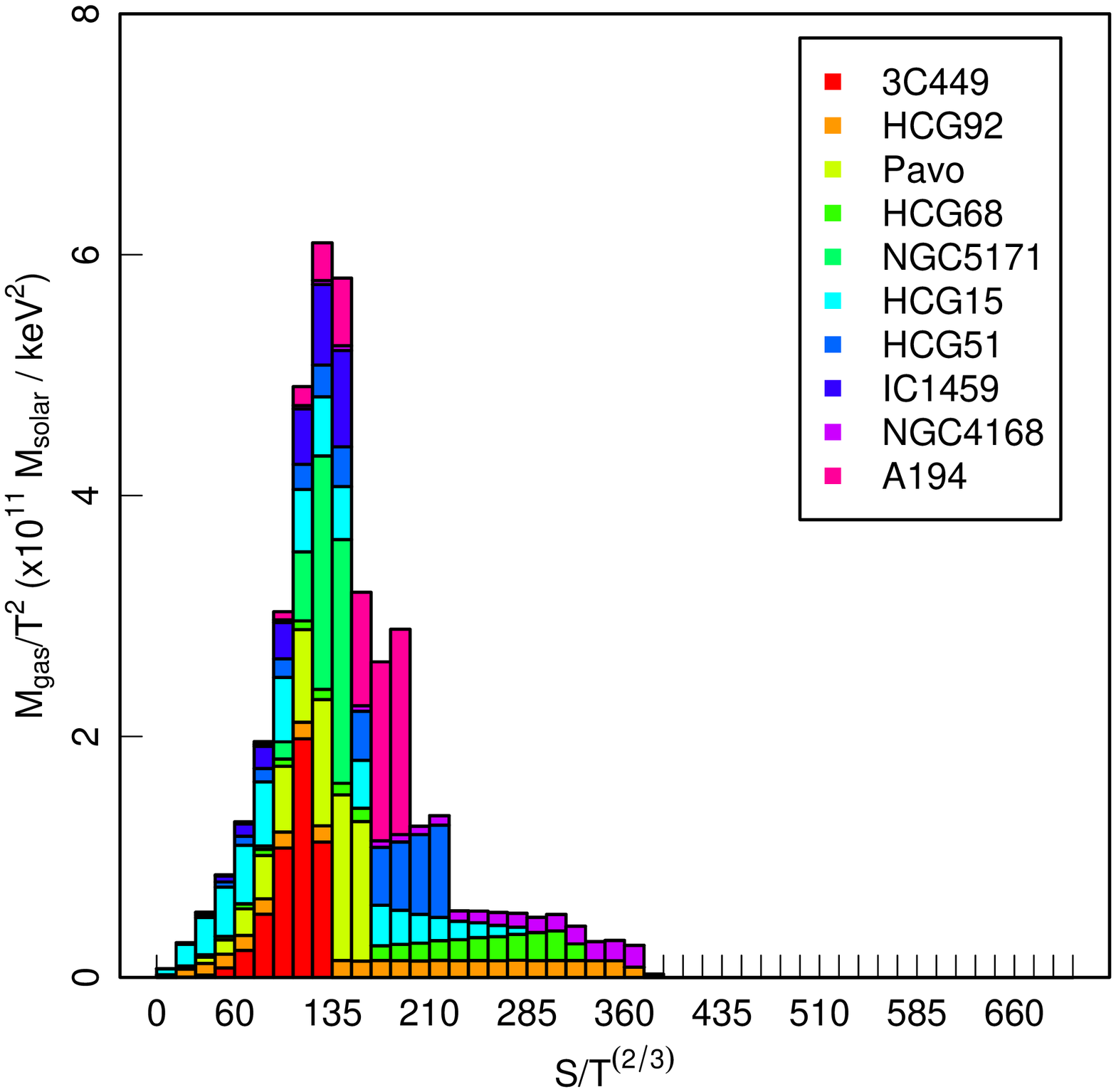}
\caption{Histogram of gas mass as a function of entropy for the NCC groups, with each group colour-coded as shown in the legend. This shows the total gas mass in a particular entropy bin, indicating where the majority of the gas mass lies. We have used the modified entropy scaling of $\bar{T}^{2/3}$ to directly compare the systems, and have scaled the gas mass by $\bar{T}^{2}$ (see text for details). }
\label{fig:Shist-NCC}
\end{figure}
Considerable insight into the heating and cooling processes affecting
the intracluster medium can be gained from looking at how the entropy
is distributed as a function of gas mass of the system
\citep[e.g.][]{voit03b}. Having fitted power-laws to the scaled
entropy profiles of all the groups individually (see Section
\ref{sec:S-ind}), we used these alongside $\beta$-model fits to the
gas density profiles to determine the gas mass in a series of radial
shells. The shells were stepped in radius from 0.0 to 0.5\,\RF, with
width 0.0005\,\RF, and we set the characteristic radius for a shell to
be its mean radius. The fits to the entropy and gas density profiles
were used to interpolate the entropy and density at the these
characteristic radii. The gas mass in each shell, \ensuremath{M_{gas}}
was determined as follows,
\begin{equation}
M_{gas}=4\pi~r^2~\rho~\delta\!r, 
\end{equation}
where \ensuremath{r} is the characteristic radius of the shell, $\rho$
is the density evaluated at radius \ensuremath{r}, and
\ensuremath{\delta\!r} is the width of each shell, i.e. 0.0005\,\RF.

With entropy and gas mass then described by the same set of radial
bins, we set up a series of scaled entropy bins of width
15\keV$^{1/3}$\,cm$^{2}$, and determined the total gas mass in each
entropy bin. To be able to compare groups directly, we scaled the gas
mass by \ensuremath{\bar{T}^2} to remove the dependence of gas mass on
the mass of the system as a whole. This is a result of the applied
entropy scaling which gives a gas density scaling $\propto$
$\bar{T}^{1/2}$ (see Section \ref{sec:density}). In a particular shell of
mass \ensuremath{M_{gas}},
\begin{equation}
M_{gas}=4\pi~r^2~\rho~\delta\!r~\propto~r^3~\rho.
\end{equation}
From the virial theorem, the radius $r$ is proportional to \ensuremath{\bar{T}~^{1/2}}, and with the density $\rho$ scaling as \ensuremath{\bar{T}~^{1/2}} we find,
\begin{equation}
M_{gas}~\propto~\bar{T}^2. 
\end{equation} 

Dividing the sample into CC and NCC systems, we can
examine the entropy distribution in both types of system. Figure
\ref{fig:Shist-CC} shows a histogram of the gas mass as a function of
entropy. This indicates where the majority of the gas lies in terms of
its entropy value, for each of the CC groups. The shape of these
histograms requires careful interpretation. Since entropy
rises outwards, the peak in these histograms is
produced as a consequence of the limiting upper radius of the gas mass
calculation (0.5\,\RF). At this radius, some groups have reached a
higher entropy level than others, which produces the high entropy
tail in the plot. It is
evident is that there are five CC systems (HCG~42, NGC~2300,
NGC~2563, NGC~5846 and SRGb119) where higher entropy values are
reached at 0.5\,\RF\ compared to the remainder of the CC
sample. NGC~2300 reaches very high entropies ($>$
600\,keV$^{1/3}$\,cm$^{2}$), although the entropy profile is
extrapolated from just 4 data points, all within 0.2\,\RF, so the
result shown here should be interpreted with caution. The CC systems
generally show a sharp increase in scaled gas mass with scaled entropy
on the leading edge of the peak, and show a dominant peak at
approximately 105--120\,keV$^{1/3}$cm$^{2}$, indicating that similar
entropy levels are reached by the groups at 0.5\RF.

The entropy distributions for the NCC systems are shown in
Figure \ref{fig:Shist-NCC}. In comparison to the CC systems, the NCC groups
show a wider diversity in entropy distributions, as can also be seen
in Figures \ref{fig:CC-theory} and \ref{fig:NCC-theory}.
This is in agreement with the observed
tight trends for the entropy profiles of CC groups, seen in
Figure \ref{fig:SprofCCNCC}, whilst the NCC systems show a
wider spread in their entropy properties. 

The entropy histograms also
show a lack of lower entropy gas in the NCC
systems --- in the CC systems the mean scaled gas mass with
scaled entropy $\leq$ 90~keV$^{1/3}$\,cm$^{2}$ is
$\sim$\,9.8$\times$10$^{10}$\,$M_{\odot}$~keV$^{-2}$, compared to
$\sim$\,5.0$\times$10$^{10}$\,$M_{\odot}$~keV$^{-2}$ in the NCC
 systems, thus confirming the results seen in Section
\ref{S:props}. Also, it is apparent that the entropy at which scaled
gas mass peaks is slightly different, with the CC systems
peaking in the bin 105--120~keV\,cm$^{2}$, whereas the NCC
histogram peaks at 120--135~keV\,cm$^{2}$.

Since this analysis employs scaled
entropy and scaled gas mass, the mean effect of system mass should have
been scaled out. We find the mean total scaled gas
mass, evaluated within 0.5\,\RF, for the CC and NCC
groups to be (4.7$\pm$0.5)$\times$10$^{11}$$M_{\odot}$\,keV$^{-2}$ and
(4.1$\pm$0.5)$\times$10$^{11}$$M_{\odot}$\,keV$^{-2}$ respectively. 
Therefore, there is no significant difference in the
overall scaled gas content of these systems within 0.5\,\RF, indicating that the
different core properties do not result from global differences in gas
content.

\subsection{Comparison to a theoretical entropy distribution}
\begin{figure*}
\includegraphics[width=15cm]{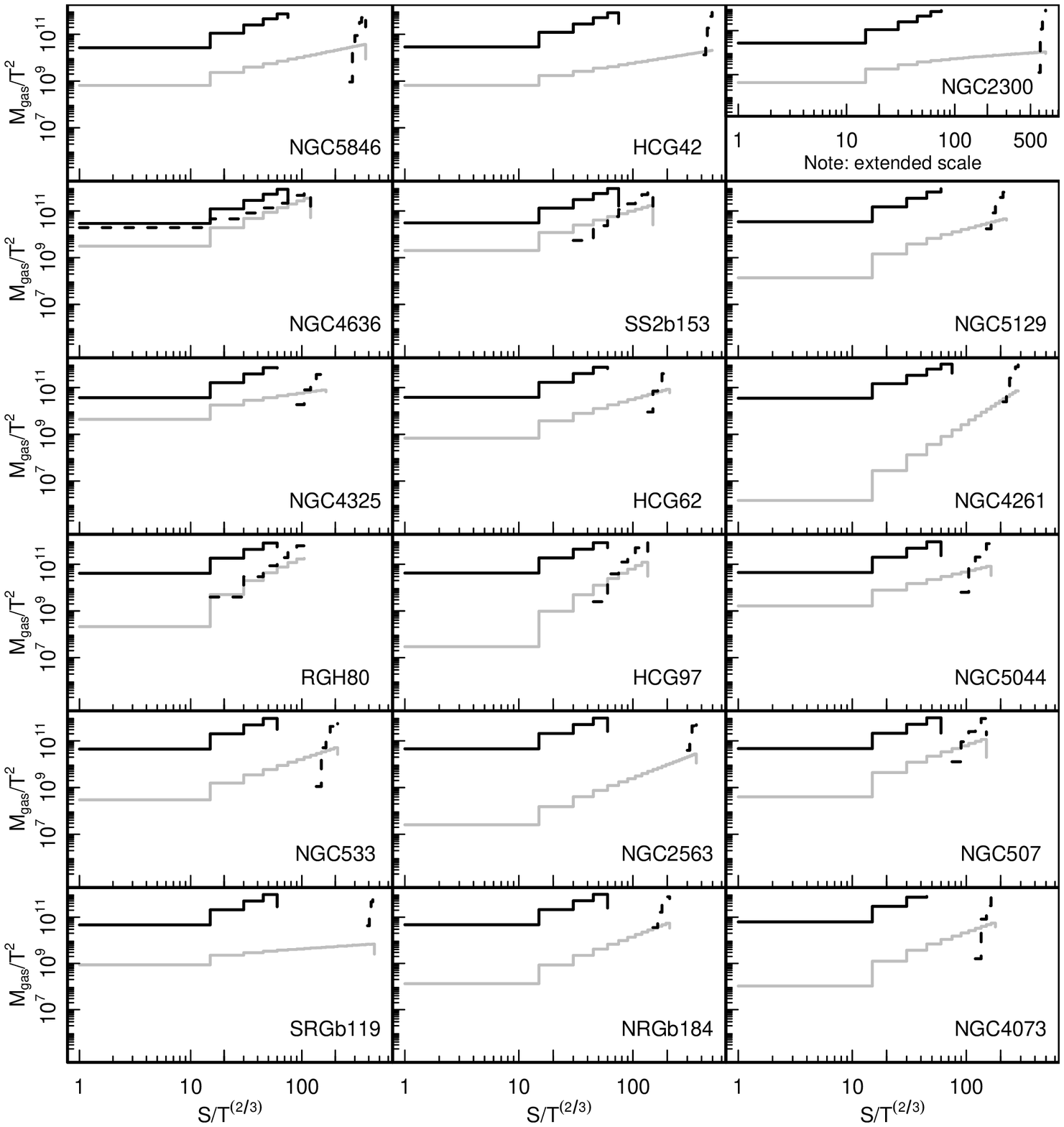}
\caption{The observed entropy distributions of the CC groups (shown in grey) 
and the calculated theoretical entropy distribution (shown in
black). In all cases the observed distribution extends to a higher
maximum entropy, compared to the theoretical distribution. The dashed
line shows the entropy distribution resulting from a modification in
the form of an entropy shift, followed by radiative cooling (see
Section \ref{sec:S-mod}). Note the extended x axis in the case of
NGC~2300, to show the high entropies reached by this group (this is
due to a significant extrapolation, see Section \ref{sec:Sdist}). The
groups are ordered in increasing mean temperature, from top left to
bottom right, and all curves extend to 0.5\,\RF.}
\label{fig:CC-theory}
\end{figure*}
\begin{figure*}
\includegraphics[width=15cm]{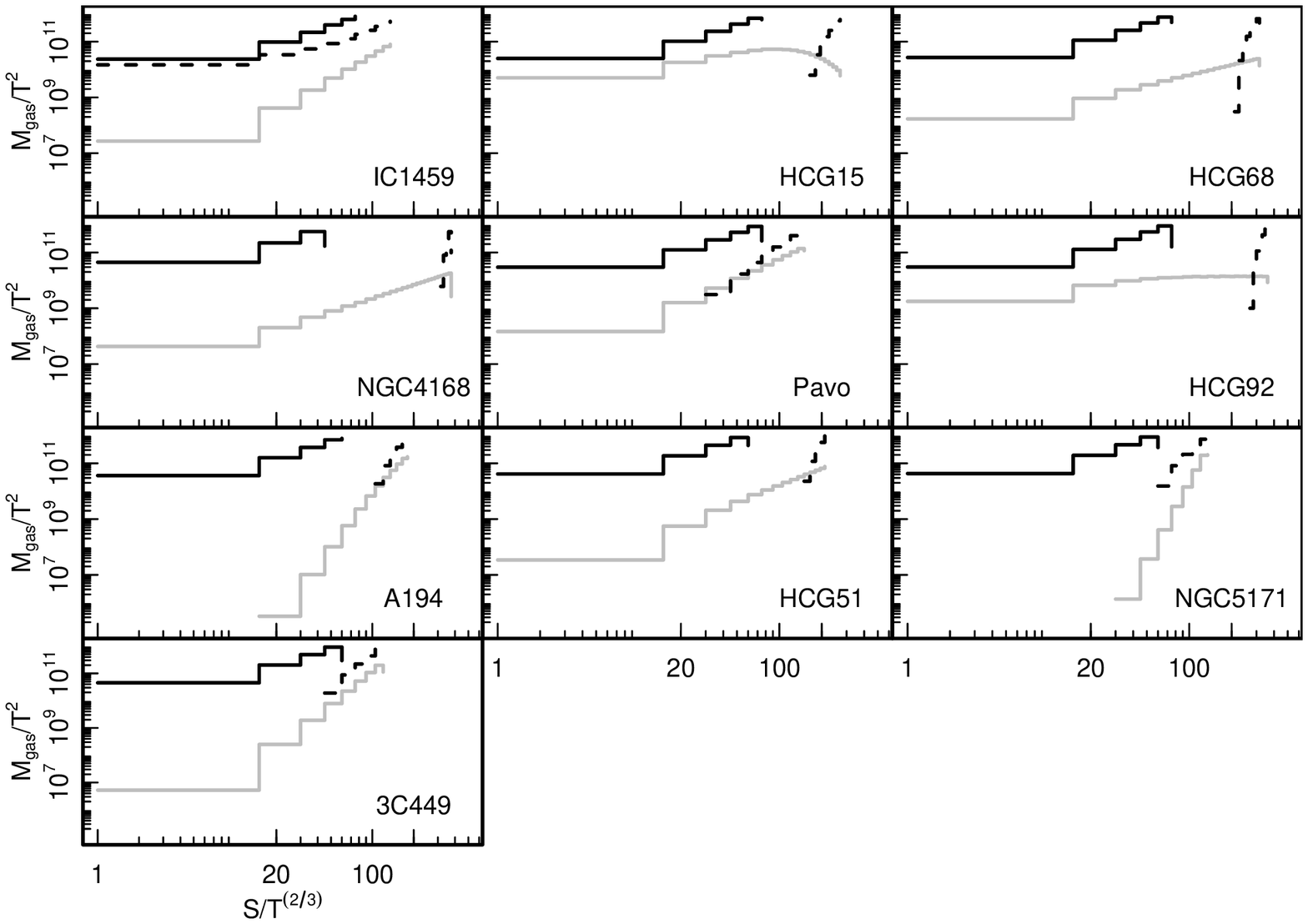}
\caption{The observed entropy distributions of the NCC groups 
(shown in grey) and the calculated theoretical entropy distribution
(shown in black). Like the CC systems, in all cases the
observed distribution extends to a higher maximum entropy compared to
the theoretical distribution. The dashed line shows the entropy
distribution resulting from a modification in the form of an entropy
shift, followed by radiative cooling (see Section
\ref{sec:S-mod}). The groups are ordered in increasing mean
temperature, from top left to bottom right, and all curves extend to
0.5\,\RF.}
\label{fig:NCC-theory}
\end{figure*}
In an effort to ascertain the magnitude of the effects of feedback
processes acting on the observed entropy distributions, we can compare
these distributions to theoretical distributions where the effects of
non-gravitational processes have been ignored. The analytical models
of \citet{voit03b} show that the radial distribution of entropy 
is approximately proportional to the enclosed gas mass if the gas 
accreted is cold and the
accretion is smooth. Looking specifically at groups, \citet{voit03b}
postulate smoother accretion in groups relative to clusters, to
explain the observed differences in the entropy profiles of systems
spanning this range in mass. We can begin to evaluate the effects of
feedback on the group gas by first computing the expected entropy
histogram for a system where only gravitational processes are
considered. The observed difference between this `theoretical'
distribution and the observed distribution can then be used as a probe
of the impact of feedback processes.

We firstly require a baseline
entropy profile $S(r)$ for a system unaffected by non-gravitational
processes. For this, we adopt the baseline entropy profile of
\citet{voit05b}, derived from smoothed particle hydrodynamics (SPH) cosmological 
simulations of galaxy clusters, without any cooling or feedback
processes. This has the form
\begin{equation}
\label{eqn:vkb}
S(r) = 1.32~S_{200}\left(\frac{r}{r_{200}}\right)^{1.1},
\end{equation}
where $S_{200}$ is defined as
\begin{equation}
S_{200} = 362~\keV~\cm^2\left(\frac{T_{200}}{T_X}\right)
\end{equation}
where we have dropped the cosmological factors from \citet{voit05b}, which
are negligible given the low redshift of our sample. For consistency within this
work, we use $S$ to denote entropy, equivalent to the $K$ used by
\citet{voit02} and \citet{voit05b}. For the temperature $T_{X}$ we use
our mean X-ray derived temperatures, and calculate $T_{200}$ in the
manner described by \citet{voit05b}. This requires knowledge of both
$r_{200}$ and the mass enclosed at this radius, $M_{200}$, which is
simply the volume at $r_{200}$ multiplied by 200\,$\rho_{crit}$. The
former can be calculated by determining a conversion factor from the
known $r_{500}$. We assume an NFW profile \citep[e.g.][]{navarro97},
evaluated at an overdensity of 500. \citet{sun08} found a
concentration at this overdensity of 4.2 for their sample of galaxy
groups observed with \Chandra\ and we proceed to use this
concentration in the following analysis. We find for an NFW dark
matter halo with this concentration, that $r_{500}$ =
0.66~$r_{200}$. This modifies Equation \ref{eqn:vkb} to the following,
\begin{equation}
\label{eqn:vkbmod}
S(r) = 1.32~S_{200}\left(\frac{0.66r}{r_{500}}\right)^{1.1}.
\end{equation}
Therefore, we use Equation \ref{eqn:vkbmod} to describe the
theoretical entropy profiles of our systems.

To calculate the gas mass, we assume, in line with
\citet{voit02}, that the gas density profile follows the dark
matter density profile, scaling the latter by a factor
$\Omega_{b}$ / $\Omega_{dm}$, assuming $\Omega_{b}$ = 0.022~h$^{-2}$
\citep{voit05b} and $\Omega_{dm}$ = 0.2. We then apply the same
analysis procedure as for the observed entropy distributions,
calculating first gas density and entropy as a function of radius,
before determining the gas mass in each bin of entropy. We use scaled
entropy and gas mass as for the observed histograms throughout our
comparisons. Figures \ref{fig:CC-theory} and \ref{fig:NCC-theory} show 
the observed
and theoretical entropy distributions for the CC and NCC
groups respectively. The rightmost bin of each distribution
should be treated with caution, since due to our imposed cut-off at
0.5\,\RF, these bins are incomplete. 

In all cases, the observed maximum entropy far exceeds the
theoretically derived maximum entropy, suggesting substantial
modification must have occurred to the theoretical distribution to
yield the observed distribution. We now explore some simple
heating and cooling prescriptions, to see to what extent they are
able to reproduce the modified entropy distributions observed.

\subsection{Entropy modifications}
\label{sec:S-mod}
\citet{voit02} consider three simple modifications to a baseline entropy distribution, 
which we summarise below:
\begin{enumerate}
\item A truncation in the entropy distribution where gas is removed if its entropy  
falls below a threshold value corresponding to a given cooling time. This
approximates the effects of cooling, where gas cools out to form stars, or is
heated by feedback from proximate cooled gas (e.g. via supernova feedback),
raising its entropy such that it convects to larger radii.
\item A shift in the entropy distribution, such that the entropy of all 
gas is increased by a certain baseline entropy, mimicing the effects of pre-heating.
\item Lowering of the whole entropy distribution due to 
radiative cooling. The entropy distribution of the gas to the 3/2 power 
(i.e. $S^{3/2}$) is reduced by the 3/2 power of the critical entropy ($S_{c}$) across 
the whole entropy distribution. Gas which drops below zero entropy as a result of
this modification is removed, as in model (i).
The form of the entropy reduction is based
on the approximation that the cooling function for gas in the group regime ($T<2$~keV)
has a temperature dependence $\Lambda(T)\propto T^{-1/2}$. 
\end{enumerate}
The critical entropy $S_{c}$ in models (i) and (iii) is some fraction of
$S_{200}$, the entropy at r$_{200}$. In fact, following the discussion
presented by \citet{voit02}, we specify,
\begin{equation}
\label{eqn:Sc}
\frac{S_c}{S_{200}}~\approx~0.164~\frac{T}{T_{200}}~\left(\frac{T}{2~\keV}\right)^{-1}~\left(\frac{\Omega_{M}}{0.33}\right)^{-2/3},
\end{equation}
calculating $S_{c}$ for each group, where $T$ is its mean
temperature.

It can immediately be seen by considering the differences between the observed (grey) and theoretical (solid black) distributions in Figures \ref{fig:CC-theory} and \ref{fig:NCC-theory} that simple truncation (model (i)) and shift (model (ii)) models are inadequate. Truncating the theoretical
distribution at a critical entropy would lead to a loss of low entropy
gas, but would not modify the shape of the distribution above the
critical entropy, making no improvement in the agreement between the
observed and theoretical distributions at high entropy, and making things
worse at low entropy (where low $S$ gas is actually observed in nearly
all groups). If instead we add a constant
(a `shift') to the theoretical entropy distribution, we simply move
this distribution in comparison to the observed distribution. This
will improve the agreement at high
entropy, but would leave the distribution devoid of any low entropy gas.

The radiative cooling modification implemented by \citet{voit02}
reduces the theoretical entropy distribution $S^{3/2}$ by
$S_{c}^{3/2}$ and removes all gas with an entropy less than zero. The
implications of such a model are clear; unlike the truncation and
shift models, low entropy gas will remain. However, it is not possible
to \textit{increase} the maximum entropy reached by the theoretical
model in this example, as whatever the value of $S_{c}$, the resulting
entropy is a reduction of the original. 
Therefore, a simple
radiative cooling model does not aid in matching the entropy profiles
at the highest entropies reached in Figures \ref{fig:CC-theory} and
\ref{fig:NCC-theory}.

It seems clear that a more sophisticated approach is required, and we
can envisage a model which combines the simple modifications described
above. To fix the high entropy behaviour, we first need a shift in
entropy, which we allow to subsequently cool via the radiative
modification to populate the low entropy end of the distribution. In
physical terms, this model uses an entropy shift to mimic early
pre-heating of the group gas, which subsequently cools through
radiative cooling. In fact, we can calculate the exact level of the
required entropy shift, $S_{shift}$, for the final maximum theoretical
entropy to match the maximum entropy achieved by the observed
distribution. Assuming the maximum observed entropy $S_{o,max}$ is
achieved by first adding a shift $S_{shift}$ to the maximum
theoretical entropy $S_{t,max}$, before subsequently cooling as
described above by an amount $S_{c}^{3/2}$, we have,
\begin{equation}
S_{o,max}^{3/2} = (S_{t,max}+S_{shift})^{(3/2)} - S_{c}^{3/2}.
\label{eqn:shift}
\end{equation}
Rearranging Equation \ref{eqn:shift}, we have
\begin{equation}
\label{eqn:shift-val}
S_{shift} = (S_{o,max}^{3/2}+S_{c}^{3/2})^{(2/3)}-S_{t,max}.
\end{equation}
Therefore, applying the shift $S_{shift}$ before cooling by
$S_{c}^{3/2}$ will match the right hand edge of the entropy
distribution. The dashed
lines in Figures \ref{fig:CC-theory} and
\ref{fig:NCC-theory} show this entropy modification to the theoretical
distribution. In the case of NGC~2300, we have extended the x axis to
show the high entropy behaviour at 0.5\,\RF\ (see Table
\ref{sec:S-ind}), a result of fitting a power-law to a profile of only
four data points, all located within 0.2\,\RF. Therefore, due to the
extensive extrapolation the entropy distribution of NGC~2300 should be
treated with caution.

Our applied ``shift + cool'' model is certainly more successful than models
(i)-(iii), but it is able to produce a
reasonable representation of the observed distribution in only a small
number of cases. In the CC sample, NGC~4636 and RGH~80
are best represented by the applied model, however, Hickson~97 and
SS2b153 also perform reasonably well. In the NCC sample,
Pavo, Abell~194, NGC~5171 and 3C~449 all show a reasonable agreement
between the observation and the applied model. However, the vast
majority of groups show marked differences between the ``shift +
cool'' model and the observed entropy distribution. In every case this
disagreement is due to a lack of low entropy gas in
the model. Conversely a higher gas mass than observed is usually present in
the highest entropy bins. One caveat with our model is that the total
gas mass within 0.5\,\RF\ remains fixed. In reality, this value
would change, due to movement of the gas as a consequence of the
applied heating and cooling mechanisms. For example, if low entropy
gas is removed from the distribution, higher entropy gas will flow
in. Boosting the entropy of the gas would in reality cause it to
expand, such that a proportion of the gas mass at high entropy
would actually fall outside 0.5\,\RF, and the applied entropy shift
will have been underestimated.  Therefore, once the expansion of the
gas is taken into account, the applied entropy shift needs to be somewhat
{\it larger} than shown in Figures \ref{fig:CC-theory} and
\ref{fig:NCC-theory}. This will exacerbate the difficulties of the model,
which arise (see discussion below) from the large entropy boost required
in many systems.

The required entropy shifts range between 115\,keV\,cm$^{2}$ (RGH~80)
and 567\,keV\,cm$^{2}$ (NGC~2300). The models which re-create the
observed entropy distribution most successfully are those where the
value of the applied entropy shift is very similar to the critical
entropy $S_{c}$ (Equation \ref{eqn:Sc}) which governs the radiative
cooling (Equation \ref{eqn:shift}), since in these cases, the shift
provides a good match at the high entropy end of the distribution, whilst
the cooling is able to repopulate the low entropy end.
For example, in NGC~4636, the
applied entropy shift is 148\,keV\,cm$^{2}$ compared to a critical
entropy of 164\,keV\,cm$^{2}$ and in RGH~80, the entropy shift is
115\,keV\,cm$^{2}$ compared to a critical entropy of
109\,keV\,cm$^{2}$.  The groups for which this works are
those with the \textit{smallest difference} in maximum entropy
between the observed and theoretical distributions. In the
majority of cases however, the required shift to match the high
entropy end is much larger than the critical entropy, in which
case the gas entropy is boosted to the point where cooling has little effect.

\subsubsection{The effects of metallicity}
\begin{figure}
\includegraphics[width=8cm]{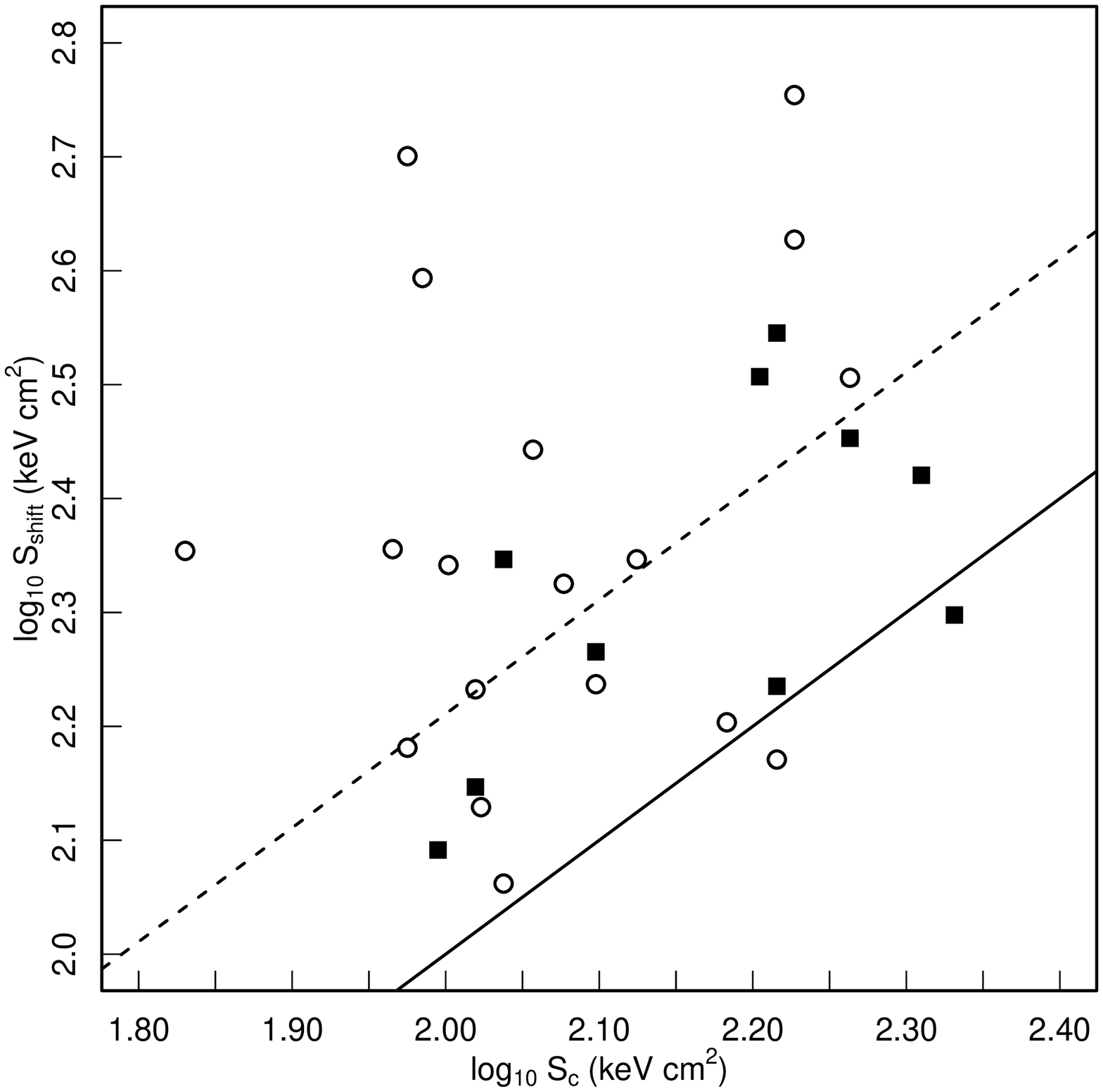}
\caption{The required entropy shift $S_{shift}$ versus the critical entropy $S_{c}$ for the CC (open circles) and NCC (filled squares) groups in the sample. The solid line is the line of equality, and the dashed line shows the same but re-normalised by the factor $\sim$1.6.}
\label{fig:KsKc}
\end{figure}

The \citet{voit02} prescription for the critical entropy assumes 
that cooling has proceeded for 15\,Gyr, and that the emissivity of the gas
corresponds to a metallicity of 1/3 solar. However, in the group regime, 
emissivity is a strong function of metallicity (due to strong line cooling)
and in CC groups there is a strong central abundance gradient 
(Rasmussen \& Ponman 2007, Johnson et al., in prep.)
leading to central metallicity close to solar values.
We can assess the potential
effects of such an abundance gradient on our `shift + cool' model by
instead assuming solar metallicity. This is assumed across the radial
range, so will in fact be an over-estimate of the effects of a
metallicity gradient. For a hot plasma at 1\,\keV, the
emissivity scales up by a factor of $\sim$\,2.0 when the metallicity
increases from 1/3 solar to solar. Equation 17 of \citet{voit02}
describes the rate of change of entropy as follows,
\begin{equation}
\frac{dS^{3/2}}{dt}~\propto~ T^{1/2}\,\Lambda(T),
\end{equation}
where our $S$ is equivalent to the \citet{voit02}
$K$. The \citet{voit02} model assumes the right hand side of this
relation to remain constant, such that it can be simply integrated,
\begin{equation}
\int{dS^{3/2}}~\propto~\int{T^{1/2}\,\Lambda(T)\,dt}.
\end{equation}
Integrating over a Hubble time ($t_{H}$) yields,
\begin{equation}
S_{c}~\propto~\left[T^{1/2}~\Lambda(T)~t_{H}\right]^{2/3},
\end{equation}
Hence the effect of a change in emissivity, caused by an
increase in metallicity, is to increase the value of $S_{c}$. The
factor increase is equal to the factor increase in emissivity, to the
power 2/3 (i.e. 2.0$^{(2/3)}$ $\approx$ 1.6). In Figure \ref{fig:KsKc}
we show the size of the required entropy shift $S_{shift}$ versus the
critical entropy $S_{c}$ in our `shift + cool' model. Since almost all groups
contain observable low entropy gas, the model can only be successful if
$S_c$ is similar to or larger than $S_{\rm shift}$. For the unmodified
cooling threshold (solid line) it can be seen that only a few groups
satisfy this requirement. However, with the factor 1.6 scaling in $S_c$,
the model could be successful for at least half the groups,
although approximately one third
of the systems still require entropy shifts much greater than the
critical entropy. 

Our model is still very simple, of course. The uniform solar abundance
will tend to overestimate the effects of cooling, as will the generous timescale
of 15\,Gyr allowed for the cooling to proceed. On the other hand, we
have not allowed for the hierarchical merger history of the systems.
This requires investigation with more sophisticated cosmological models.
Nonetheless, we tentatively conclude that the magnitude of the entropy
shift required in a significant fraction of groups is so high that
our `shift + cool' model will be unable to reproduce their properties.
\subsection{Effects of feedback}
\label{sec:fback}
\begin{figure*}
\includegraphics[width=15cm]{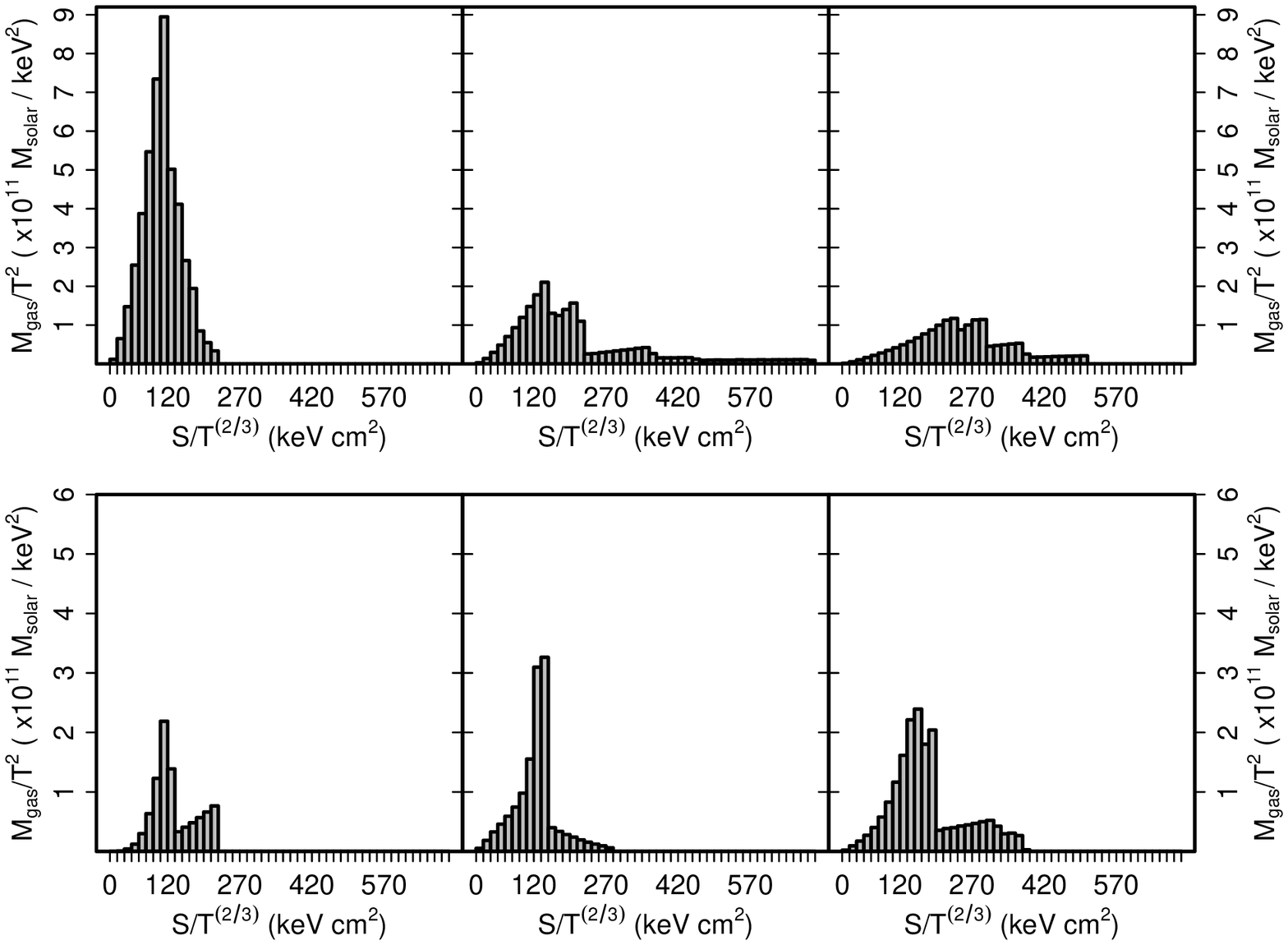}
\caption{The entropy distribution for stacked CC groups (top row) and stacked NCC groups (bottom row), in three bins of feedback impact, $f_{grp}$ (see text for definition). Feedback impact is increasing from left to right in both rows, from 0.0 $<$ $f_{grp}$ $\leq$ 0.22 to 0.22 $<$ $f_{grp}$ $\leq$ 0.6 and the rightmost panels show $f_{grp}$ $>$ 0.6. As the impact of feedback increases, so to does the maximum entropy reached by the groups, both for CC and NCC systems.}
\label{fig:Shist-fb}
\end{figure*}
We turn now to the link between modification of the entropy distribution
and the feedback processes which are most likely responsible for
this modification.
If we make the assumption that the most prominent source of feedback
in galaxy groups comes from the member galaxies, either through
energy input from supernovae or feedback from active galactic nuclei
(AGN), we can begin to quantify the likely impact of feedback. We construct
a rough measure of `feedback impact' using the ratio of the optical
luminosity of the group to the thermal energy of the gas: 
$f_{grp}=L_{B,group}/(M_{gas}T_{x})$, where $T_{x}$ is the mean
X-ray temperature of the system, and \ensuremath{M_{gas}} is the total
gas mass at 0.5\,\RF\ (derived as described in Section~\ref{sec:Sdist}).
The motivation here is that integrated feedback from supernovae 
should scale approximately with the stellar mass, and given the 
relationship \citep{magorrian98} between the mass of supermassive black holes
and the stellar mass of the spheroids they inhabit, integrated AGN
feedback might also be expected to scale roughly with stellar mass.
The effect of a given amount of injected energy on the intergalactic
gas will depend on the factor by which it increases the existing
total thermal energy, which motivates the denominator, $(M_{gas}T_{x})$,
in our expression for $f_{grp}$. This measure of feedback impact therefore has units of $L_{B,\odot}$\,\Msol$^{-1}$\,\keV$^{-1}$. For clarity we will not show the units in the remainder of this section.

Our measure of feedback impact is admittedly imperfect. In the first
place, the  $K$-band luminosity would provide a better measure
of stellar mass, since the $B$-band is strongly affected by young stellar
populations. A full stellar population analysis would be even better,
of course. However, we have consistent measurements of $B$-band luminosities
for all the groups in our sample, and are encouraged to believe that biases
due to young stars will not have a major impact, by
the fact that X-ray bright groups have a low spiral
fraction \citep{helsdon03b}. A second respect in which our formula
for $f_{grp}$ is imperfect is that the denominator actually reflects
the thermal energy of the gas {\it after} the feedback has had its
effects, rather than before. We explore the effects of modifying the
denominator at the end of this section, and find that it makes no 
qualitative difference to our conclusions.

Assuming that $f_{grp}$ is monotonically related to
feedback impact, we now use it to divide our
sample into 3 subsets of (almost) equal size, covering ranges in feedback 
impact of 
0.0 $<$ $f_{grp}$ $\leq$ 0.22, 0.22 $<$ $f_{grp}$ $\leq$ 0.6 and 
$f_{grp}$ $>$ 0.6. These bins contain 10 (8 CC + 2 NCC), 
9 (6 CC + 3 NCC) and 9 (4CC + 5 NCC) groups respectively. 
 Figure \ref{fig:Shist-fb} shows the stacked
entropy distributions for the CC (top row) and NCC (bottom row)
groups, shown in order of increasing feedback impact from left to
right. In both samples, we see a trend for an increase in the amount
of gas at higher entropies as the feedback impact increases. There is
also a suggestion of the peak of the distribution moving to higher
entropies as the level of feedback impact increases. Qualitatively,
this appears to indicate that the gas in groups with the largest
capacity for feedback from their galaxy members (in the form of energy
deposited into the intracluster medium from supernovae explosions or
AGN) has been pushed to higher entropies. This is true regardless of
whether the system has a CC or not. 

Given the similarity of the entropy distributions in the
highest feedback bins, we choose to combine these bins. Therefore, the
sample now consists of a `lower' feedback bin ($f_{grp}$ $\leq$ 0.22)
and a `higher' feedback bin ($f_{grp}$ $>$ 0.22). Table \ref{tab:fgrp}
shows the break-down of our groups into the CC/NCC and high and
low feedback classes. In brackets we show the {\it expected} number of
groups in each of the four categories, given the relative ratios of
CC to NCC groups, and high to low feedback systems (18 to 10, in each
case) if the two classifications (CC/NCC and high/low feedback) were
completely independent. This suggests that there is some tendency for the
NCC systems to have a higher feedback impact than CC systems. The
median feedback impact for the whole sample is 0.31, whereas for CC
systems it is 0.27 and for NCC systems 0.62.  We quantify
the significance of the relationship between feedback and CC status by
calculating how likely it is that there should be at least 8 groups in
both the high feedback, NCC class, and the low feedback, CC class, given
that only 6.4 systems were expected in each, using the binomial
probability distribution. This probability is 6 percent, so we
conclude that there is some evidence for a relationship between the
two properties, but it is clear from Table \ref{tab:fgrp} that the
relationship is rather weak.

\begin{table}
 \centering 
 \caption{The distribution of CC and NCC systems between the higher and lower feedback bins. The numbers in brackets show the expected number of groups in each of the four categories, if the classifications of CC/NCC and high/low feedback are independent.}
 \label{tab:fgrp}
 \begin{tabular}{cccc}
 \hline
 & CC & NCC & Total\\
 \hline
 High Feedback ($f_{grp}$ $\geq$ 0.22) & 10 (11.6) & 8 (6.4) & 18\\
 Low Feedback ($f_{grp}$ $<$ 0.22) & 8 (6.4) & 2 (3.6) & 10\\
 \hline
 Total & 18 & 10 &\\
  \end{tabular}
\end{table}

\begin{figure}
\includegraphics[width=8cm]{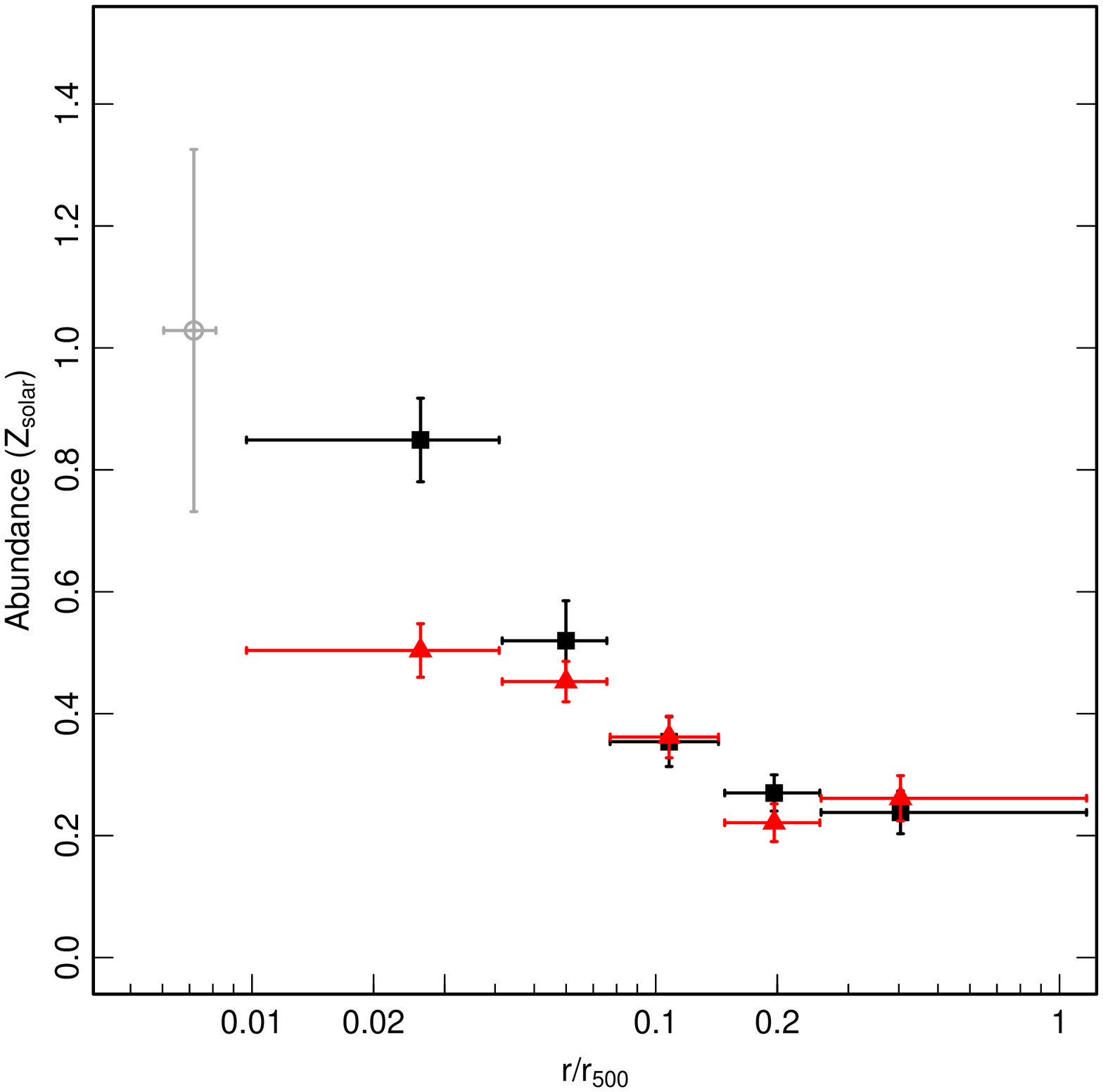}
\caption{The stacked abundance profiles of the groups with the lowest level of feedback impact ($f_{grp}$ $<$ 0.22, see Section \ref{sec:fback}), shown as black filled squares and the groups with higher levels of feedback ($f_{grp}$ $\geq$ 0.22), shown as red filled triangles. Each sample has been divided into equal bins of radius. Vertical error bars show the standard error in each abundance bin, and horizontal error bars show the width of each radial bin. The radial coverage of the low feedback systems extends to a smaller radius than the high feedback systems, and we show these remaining low feedback systems as a single grey open circle point.}
\label{fig:mmass}
\end{figure}

In addition to affecting the energy of the gas in the intracluster
medium, feedback processes can also inject and redistribute metals in the
intracluster medium. Supernova feedback will introduce more metals
into the intracluster medium, whilst AGN will not, so by examining the 
metal content of the gas in high and low feedback systems, we may hope 
to differentiate between the two sources of feedback.
Figure \ref{fig:mmass} shows the
metallicity profile of the lowest feedback ($f_{grp}$ $<$ 0.22)
systems and the higher feedback systems ($f_{grp}$ $\geq$ 0.22). We 
have stacked the abundance profiles of the two feedback sub-samples 
into equal radial bins. The lowest feedback systems extend to a smaller 
radius than the higher feedback systems. We therefore show the contribution
at small radii from the lowest feedback systems as a single grey point.
Over the majority of the profile, the low and high feedback systems
are remarkably similar. Only in the innermost radial bin do the
profiles differ significantly, with the higher feedback systems showing 
\textit{lower} central metallicities than the lowest feedback systems.
We estimated an approximate integrated metal fraction for each group by
determining the product of metallicity and gas mass, summed over a series
of radial shells, and dividing by the total gas mass. We restricted 
this calculation 
to within 0.3\,\RF, to ensure that the majority of groups have
the appropriate radial spectral coverage to avoid significant extrapolation of
the metallicity profile. For the low feedback systems, we find the mean
metal fraction (with standard error) to be 0.23$\pm$0.04, whereas the
mean metal fraction (with standard error) of the high feedback systems
is 0.25$\pm$0.03. Hence, across this radial range, the integrated 
metal fraction of the two subsets is essentially identical. This is the case
despite the abundance difference in the innermost radial bin
since the metal fraction is dominated by shells at larger radii,
where the majority of the gas lies. These results run counter
to naive expectations. We would expect higher feedback
systems to have experienced more supernova feedback, and therefore to show
higher abundances and integrated metal mass fractions.

\begin{figure}
\includegraphics[width=8cm]{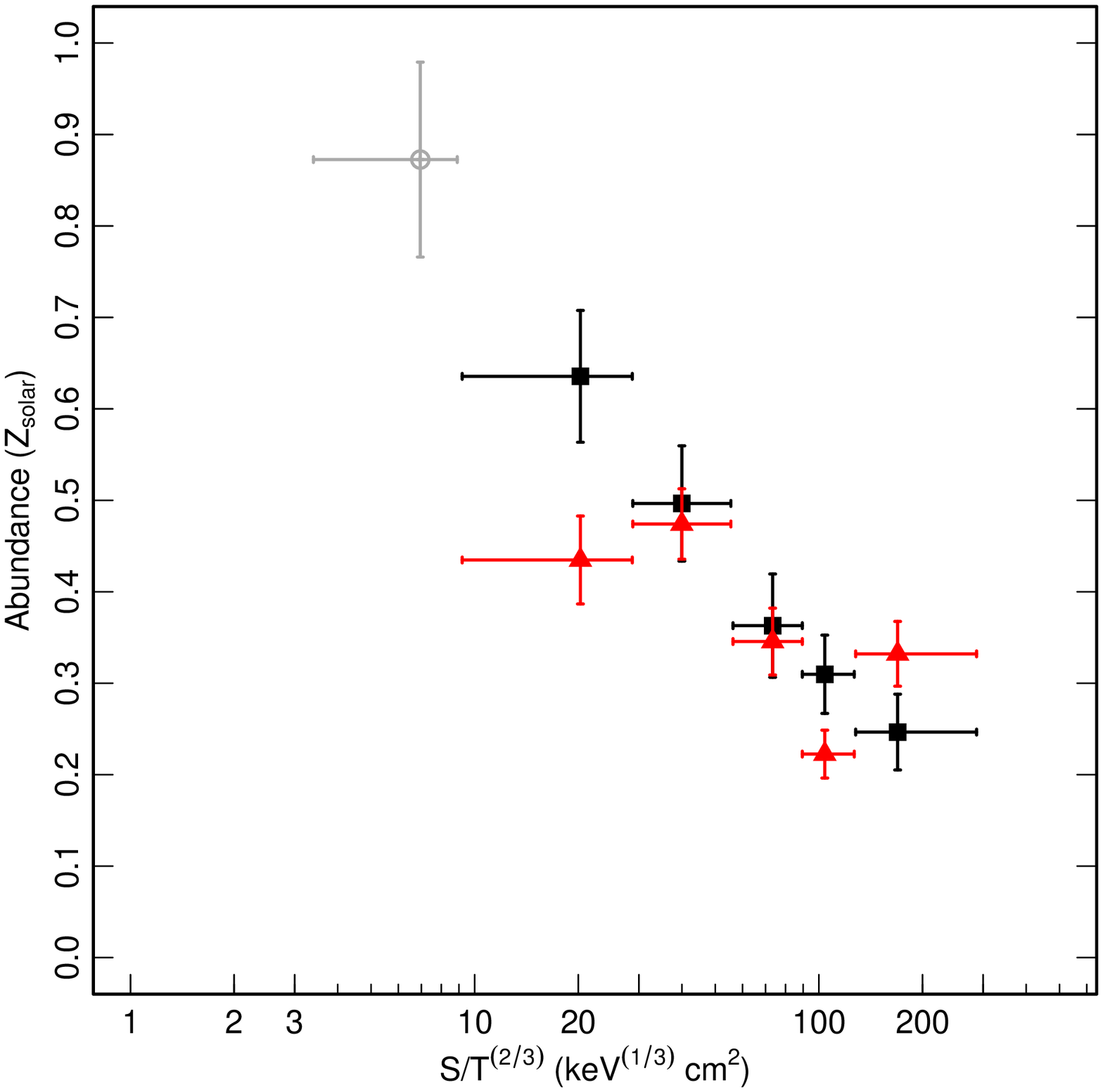}
\caption{Stacked abundance for the low feedback (black filled squares) and the high feedback (red filled triangles) systems shown in Figure \ref{fig:mmass}, but now as a function of scaled entropy. Each sample has been divided into five equal bins in scaled entropy. Vertical error bars show the standard error in each abundance bin, and horizontal error bars show the width of each entropy bin. The lowest feedback systems extend to a lower entropy, and we show these remaining low feedback systems as a single grey open circle point.}
\label{fig:mmass-e}
\end{figure}

The fact that we see no
increase in the integrated metal mass fraction in the highest feedback 
systems, 
coupled with the lower central metallicity of these groups, suggests that
AGN may provide the dominant source of feedback, as supernova feedback 
would inject metals into the gas, along with the deposited energy, leading
to extra enrichment. We have already seen
in Figure \ref{fig:Shist-fb}, that there is a lack of low entropy gas
in the high feedback systems, compared to the low feedback groups.
We can examine whether high metallicity, low entropy gas has
been boosted in entropy (generating high entropy gas with high metal 
abundance) by considering the distribution of metals as a function
of entropy, rather than radius. In Figure \ref{fig:mmass-e}, we
show the stacked abundance as a function of entropy for the low
($f_{grp}$ $\leq$ 0.22) and high ($f_{grp}$ $>$ 0.22) feedback
systems. The samples have been divided into the same bins in entropy
for comparison. As the low feedback systems extend to lower entropy,
we show the remaining data points outside the range of comparison as
an open grey circle in Figure \ref{fig:mmass-e}. In the highest
entropy gas, the gas in the low and high feedback systems is
similarly enriched. The lowest entropy gas in the highest feedback
systems is of lower metallicity, compared to gas at the same entropy
in the lowest feedback systems. Hence, if in the high
feedback systems, the high metallicity, low entropy gas has been boosted
in entropy, it would need to be pushed beyond 0.5\,\RF\ for us not
to detect it in our combined data.

One concern in studying the relationship between our feedback impact
parameter and gas entropy is that both are related to gas density. Systems
with low gas density will have high entropy, and will also (via the entry
of $M_{gas}$ into our expression for $f_{grp}$) tend to have high
feedback impact. As mentioned earlier, the denominator of the
expression for $f_{grp}$ should really be the total thermal energy
in the gas {\it before} the action of feedback, rather than afterwards.
Assuming that all systems would contain a cosmic ratio of gas to dark
matter, in the absence of cooling and feedback, we can construct an alternative
measure of feedback impact, as
\ensuremath{L_{B,group}/(M_{tot}T_{x})}. For self-similar systems at a given
epoch, the total mass and characteristic temperature are related by
$M_{tot}$ $\propto$ $T^{3/2}$. We therefore explore a modified 
definition of feedback impact of \ensuremath{L_{B,group} / T_{x}^{\,5/2}}. 

Using this revised parameter to partition our group sample into
high and low feedback classes, we find results very similar to those
obtained with our earlier definition of feedback impact: gas with
entropy greater than $\sim$30\keV$^{(1/3)}$\,cm$^{2}$ is similarly
enriched in the low and high feedback systems, and in the innermost
entropy bin, the highest feedback systems reach lower central
metallicity. We also find a trend for the higher feedback systems to reach
higher entropies than the lowest feedback groups, as seen in Figure 
\ref{fig:Shist-fb} for our unmodified measure of feedback impact. We conclude 
that the gas density dependence of our original 
feedback measure does not affect the results.


\subsection{The \ensuremath{\sigma-T} relation}
\begin{figure}
\includegraphics[width=8cm]{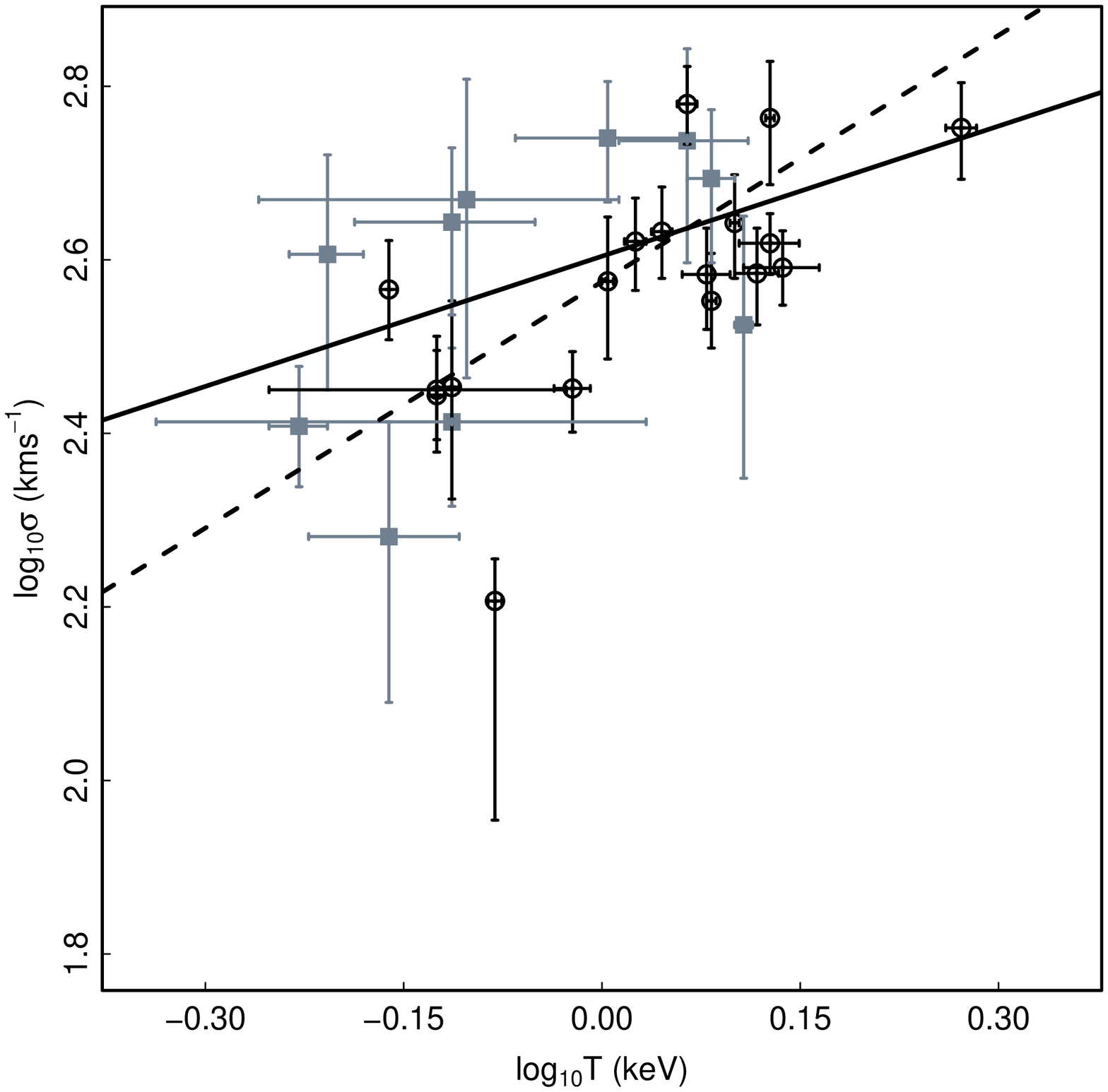}
\caption{The $\sigma$-$T$ relation for the groups in the sample. Black open circles denote the CC groups, and grey filled squares denote the NCC systems. The solid line shows the relation $\beta_{spec}$ = 1, and the dashed line shows a BCES orthogonal regression fit to the data (see text for details).}
\label{fig:global}
\end{figure}
We can further investigate the impact of feedback by considering the
quantity $\beta_{spec}$, which is the ratio of the specific energy of
the galaxies to the specific energy of the gas,
\begin{equation}
\beta_{spec}=\frac{\mu~m_{p}\sigma^2}{kT} ,
\end{equation}
where \ensuremath{\mu} is the mean mass per particle,
\ensuremath{m_{p}} is the proton mass, \ensuremath{k} is
Boltzmann's constant, and $T$ is the temperature of the gas.
If feedback raises the energy of the gas, then the consequence
is to reduce $\beta_{spec}$, as the subsequent specific energy of the
gas would dominate over the specific energy of the galaxies.

We can use the \ensuremath{\sigma-T} relation to investigate where
groups lie with respect to $\beta_{spec}$ = 1. From a theoretical perspective
we expect to see similar specific energy in gas and galaxies, unless
feedback has had a major effect, and in the case of rich clusters,
observations have shown that relaxed systems generally lie along
a trend in the $\sigma - T$ plane of the form
$\sigma$~$\propto$~$T^{0.5}$, whilst groups appear to follow a steeper
trend (see the discussion in \citet{osmond04}).
We have used a BCES orthogonal regression
\citep{akritas96} to fit a straight line to the \ensuremath{\sigma-T}
relation for our group sample, yielding the fit:
\begin{equation}
log_{10}~\sigma = 0.9~(\pm0.3)~log_{10}~T+2.57~(\pm0.02).
\end{equation}
This method accounts for both the measurement errors and the unknown intrinsic 
scatter in the relation. 

Figure \ref{fig:global} shows the \ensuremath{\sigma-T} relation for
the group sample, where we show the line corresponding to
\ensuremath{\beta_{spec}=1} as a solid line. Six of the ten NCC
systems are seen to lie above the line of
\ensuremath{\beta_{spec}=1}, whereas the majority of CC systems
lie beneath this line. If the lack of central cooling in NCC systems 
resulted from strong feedback, as in the model of \citet{mccarthy08}, who invoke
pre-heating prior to cluster collapse, then the specific energy of the gas
should be higher than that of the galaxies in NCC groups,
i.e. $\beta_{spec}$ $<$ 1, whereas in practice, we see the exact opposite.
This may indicate the importance of cluster merging in the
formation of NCC systems, as in merging systems, the velocity
dispersion can be boosted, which would in turn increase
$\beta_{spec}$. This is supported by the NCC group with the highest
$\beta_{spec}$, Hickson~92 (Stephan's Quintet), which is clearly a system
currently undergoing multiple mergers \citep[e.g.][]{trinchieri03}.
\subsection{Discussion}
\label{sec:discuss}
We can connect the evidence of Section \ref{sec:Sdist} to build a
coherent picture relating galaxy group properties, entropy and the
effects of feedback. Section \ref{sec:S-mod} shows that modifications
must have been made to the gas entropy distribution in these groups. 
Comparing with a theoretical
distribution, which does not allow for any non-gravitational
processes, we find a higher maximum entropy in the observed entropy
distributions. Simple entropy modifications are not sufficient to
convert the theoretical distribution to the observed distribution, and
we rule out simple entropy shift, truncation and radiative cooling
models. Our model which mimics pre-heating followed by a bout of
radiative cooling (our so-called `shift + cool' model) only performs
well in cases where the required entropy shift is modest. In many
cases, the required large entropy shift suppresses
cooling, making it impossible to populate the low entropy end of the
observed distribution. 

To avoid this problem, we require a model which applies a
larger entropy shift to higher entropy gas. Two
physical mechanisms which might achieve this are entropy amplification
\citep{ponman03,voit03b} or episodic heating. In the former, when
pre-heated gas hits an accretion shock there is an entropy boost, and
higher entropy gas hitting the accretion shock gets a bigger boost
than lower entropy gas. In the case of episodic heating, small bursts
of heating could raise the group entropy, and subsequent cooling
operates more efficiently in the lower entropy gas. Hence, over a number of
cycles, the highest entropy gas could be boosted more than the lower
entropy gas, as required to model our observed entropy
distributions. The former mechanism could require up to two orders of magnitude
less energy than the latter, depending on when the pre-heating takes place 
\citep{mccarthy08}.

Defining feedback impact as \ensuremath{L_{B,group}/(M_{gas}T_{x})},
we find gas of higher entropy in systems with higher feedback
impact. However, there is no significant increase in metallicity in 
the highest feedback impact systems, and the metal mass fraction 
in the gas is similar to the lowest feedback systems. As
our measure of feedback impact scales with the optical luminosity of
the group, we would expect an increase in the number of supernovae,
and hence an increase in the metallicity of the highest feedback
systems. The lack of such a trend suggests that metals must have been
removed from the gas within 0.5\,\RF\, either by ejecting it to larger
radii, or by locking up a larger fraction in stars or stellar remnants, and
strongly favours AGN, rather than supernovae explosions,
as the dominant source of feedback. 

We see higher metallicity gas in the centres of the lowest feedback
groups, but across the majority of the radial range, the metallicity 
profiles of the low and high feedback systems are comparable. The
conclusion of \citet{jetha07} that central AGN within groups
influence the gas properties
of groups on a local, rather than a global, scale is supported by this
observation, as the effect of feedback is noticeable only at the
innermost radius. This contrasts with the results of \citet{croston05}, who 
found evidence for radio sources affecting the global properties of the
group gas, as represented by their location in the luminosity-temperature
plane.

In the highest entropy gas we study, whose entropy exceeds that expected
due to gravitational collapse alone, similar
levels of enrichment are reached in high and low feedback syatems. 
This again points to an AGN origin for the bulk of the feedback, though
this feedback may well have taken place within precursor structures, rather
than being driven by a central AGN within the assembled group.

Eight of the ten NCC systems have feedback impact $>$0.22, 
putting them in our highest feedback bin, and five of these have
a value of $\beta_{spec}$ $>$ 1. This is puzzling, since high feedback
impact should indicate raise the specific energy of the gas, resulting in
$\beta_{spec}$ $<$ 1. The observed  high $\beta_{spec}$ values could come
about through merging, which may therefore be the most important mechanism 
in the
formation of NCC systems. It seems that there is a complex interplay
between the gas and group properties in these systems.

\section{Conclusions}
\label{sec:conclude}
We have constructed a sample of 28 galaxy groups from the
Two-Dimensional \XMM\ Group Survey \citep{finoguenov06,finoguenov07}
and the RASSCALS groups of \citet{mahdavi05}. We have statistically
analysed the results of a novel two-dimensional spectral analysis
technique applied by
\citet{mahdavi05}, \citet{finoguenov06} and \citet{finoguenov07} to
high quality \XMM\ data. This is the largest group sample that has
been analysed in a consistent manner with \XMM\ data, and the size of
the sample allowed a division on the basis of the presence of a cool
core. We find 18 groups to exhibit cool cores (CCs) and 10 to be classified
as non cool cores (NCCs). This latter sample is the first of its kind in the
group regime, and the analysis of these groups provides useful
insights into the variation in the observed properties of these
systems. We summarise our main results below:
\begin{enumerate}
\item We have measured the ratio of the central temperature drop (defined
between the peak temperature and that at 0.01\,\RF) to the peak
temperature in the CC systems. This fractional central decline in
$T$ is smaller than in clusters. Within our whole group sample, 
we find no significant trend with mean temperature,
however, the hottest system in our sample (NGC 4073, $\bar{T}$ = 1.87~keV) 
has an unusually small temperature decline, and if this group is excluded
we find a positive correlation between CC strength and $\bar{T}$
which is significant at the $>$ 1$\sigma$ level.
\item We find radially smaller CCs in the coolest systems, 
confirming the results of \citet{rasmussen07}, and find no CC 
systems with temperatures less than $\sim$0.7\keV, indicating a possible 
lack of CC systems at lower temperatures.
\item We have investigated the relationship between temperature and 
entropy evaluated at 0.1\,\RF, incorporating the cluster sample of 
\citet{sanderson08}. Applying a BCES orthogonal regression \citep{akritas96} 
yields a slope of 0.79$\pm$0.06, in agreement with the \Chandra\ sample of 
\citet{sun08}. There is a large amount of scatter in the 
group data, and the fit is mostly constrained by the addition of the 
cluster points, leading us to adopt the modified entropy scaling 
\citep[$S \propto T^{2/3}$ of][]{ponman03} for purposes
of scaling group properties in this paper.
\item The entropy profiles of cool groups (defined as those with
mean temperature below the median value for our sample of 1.035~keV)
are significantly steeper than those of hotter groups. The slope of the
entropy profiles of the Cool subsample is found to be 0.77$\pm$0.03,
compared to 0.65$\pm$0.02 for the Warm group subsample.
\item Comparing the entropy profiles of CC and NCC groups,
we find there to be more scatter in the entropy profiles for the latter,
which is consistent with the larger scatter shown in
their temperature profiles. The central entropy, within $\sim$\,0.1\,\RF,
is also flatter for the NCC systems, whereas in CCs, 
the entropy profiles decrease steadily into the centre of the
system. There is no evidence for a central entropy
pedestal, as required for clusters \citep{donahue06}, in the CC groups.
\item Examining the pressure and density profiles, we find
a marked difference in the pressure profiles of Cool and Warm groups,
with the scaled pressure profile of the former rising to higher central
values than in warmer systems. We attribute this difference to the
increased dominance of the stellar mass of the brightest group galaxy
in the centres of the cooler systems. Comparing CC and NCC systems, 
we see slightly steeper density and pressure profiles within
the cores of CC systems, which is attributable to their central temperature 
decline. Outside the typical radius of the CC
($\sim$0.1\,\RF), we find the two subsamples to have similar gas profiles.
\item We have investigated the entropy distribution of the groups by
plotting the observed gas mass in bins of scaled entropy. 
The total gas mass
within 0.5\,\RF\ is comparable for the CC
(4.7$\pm$0.5$\times$10$^{11}$M$_{\odot}$~keV$^{-2}$) and NCC
groups (4.1$\pm$0.5$\times$10$^{11}$M$_{\odot}$~keV$^{-2}$). 
Comparing the observed entropy distributions to the theoretical expectation if
non-gravitational processes are ignored, and find that the observed
distributions generally reach much higher entropies within 0.5\,\RF. In trying
to reconcile the differences between the observed and theoretical
histograms, we find that simple modifications (shift, truncation or
radiative cooling) are not sufficient to bring the theoretical
distribution in line with that observed. A `shift + cool' model which
aims to match the high entropy behaviour of the two distributions
performs well when the entropy shift is small, but in many
cases the large entropy boost suppresses cooling, resulting in an
entropy profile which lacks the low entropy gas required by
the observations. This suggests a process whereby the entropy shift
is larger for higher entropy gas. Potential mechanisms for achieving
this are either through entropy amplification or episodic heating. We note
that the former is likely to be more energy efficient.
\item We define the `feedback impact' of a group, and find that systems
with the highest feedback impact reach higher scaled entropies within
0.5\,\RF, regardless of whether or not they have CCs. We find
higher metallicity in the central regions of the lowest feedback
systems, but over the majority of the radial range, the metallicity profiles
of the low and high feedback systems is comparable.
If low entropy
metal-rich gas in the highest feedback systems has been boosted in
entropy, it has been pushed outside 0.5\,\RF. There is no evidence for
an increase in the metal content with the level of feedback impact,
which leads us to favour AGN, probably acting before group assembly, 
as the dominant source of feedback, rather than
supernova explosions. We test for bias in our definition of feedback
impact by scaling by the total mass of the system (approximated as
$\propto$ $T^{3/2}$) rather than the gas mass. Changing our
definition of feedback impact in this way does not affect our results. 
\item Fitting the $\sigma$ - $T$ relation to the group sample yields a 
steep slope
of 0.9$\pm$0.3. Six of the NCC groups lie above the
$\beta_{spec}$=1 line, which seems inconsistent with models which invoke
feedback to eliminate CCs in these systems. We suggest that
group-group mergers are more likely responsible for the elimination
of CCs.
\end{enumerate}
\section*{Acknowledgments}
We would like to thank for the anonymous referee for their very helpful 
comments. We thank Jesper Rasmussen for his very helpful suggestions and
comments on the original manuscript, and Ian McCarthy for valuable
discussions and for drawing our attention to the potentially important
effects of metallicity gradients on the cooling model. We 
would like to thank Alastair Sanderson
for providing the entropy data from his cluster sample, and for a number
of useful discussions and suggestions. RJ acknowledges
support from STFC/PPARC and the University of Birmingham. AF acknowledges 
support from BMBF/DLR under grants 50OR0207 and 50OR0204 to
MPE. AF thanks the University of Birmingham for hospitality during his
frequent visits. This research has made use of the NASA/IPAC Extragalactic 
Database (NED)
which is operated by the Jet Propulsion Laboratory, California
Institute of Technology, under contract with the National Aeronautics
and Space Administration.
\bibliography{/data/ria/latex/ria_bibtex}
\label{lastpage}

\end{document}